\documentclass[paper,11pt]{JHEP3}
\usepackage[centertags]{amsmath}
\usepackage{amsfonts} 
\usepackage{amssymb}

\newcommand{\cA}{{\cal A}}

\newcommand{\cC}{{\cal C}}

\newcommand{\cE}{{\cal E}}
\newcommand{\cF}{{\cal F}}
\newcommand{\cG}{{\cal G}}

\newcommand{\cN}{{\cal N}}
\newcommand{\cS}{{\cal S}}
\newcommand{\cW}{{\cal W}}
\newcommand{\cV}{{\cal V}}

\newcommand{\cFpi}{{\cal F}_{\pi}}

\newcommand{\bb}{{\mathfrak b}}
\newcommand{\ff}{{\mathfrak f}}

\newcommand{\hF}{{\hat F}}
\newcommand{\hL}{{\hat L}}

\newcommand{\un}[1]{ { { \underline{#1} } } }

\newcommand{\F}{\mathbb{F}}

\newcommand{\R}{\mathbb{R}}

\newcommand{\C}{\mathbb{C}}
\newcommand{\Hp}{\mathbb{H}}

\newcommand{\uno}{\mathbb{I}}

\newcommand{\COMMENTO}[1]{}
\newcommand{\COMMENTOO}[1]{}
\newcommand{\spazio}{~~}

\newcommand{\nzm}{\, {n z m}}
\newcommand{\ce}{\oh R^2(\nu)}

\newcommand{\zlz}{{z_{L 0}}}
\newcommand{\zrz}{z_{R 0}}
\newcommand{\bzlz}{{\bar z}_{L 0}}
\newcommand{\bzrz}{{\bar z}_{R 0}}

\newcommand{\Zln}{{Z_{L \nzm}}}

\newcommand{\bZln}{{\bar Z}_{L \nzm}}

\newcommand{\Xln}{{X_{L \nzm}}}
\newcommand{\Xrn}{{X_{R \nzm}}}

\newcommand{\sqa}{\sqrt{\alpha'}}
\newcommand{\Si}{\theta}
\newcommand{\oh}{\frac{1}{2}}
\title{
Strings 
in an arbitrary constant magnetic field with arbitrary constant metric
and  stringy  form factors 
}

\author{\parbox{11.5cm}{Igor Pesando$^1$}
\\
~\\
~\\
$^1$Dipartimento di Fisica Teorica, Universit\`a di Torino\\
and I.N.F.N. - sezione di Torino \\
Via P. Giuria 1, I-10125 Torino, Italy\\
\vspace{0.3cm}
\email{ipesando@to.infn.it}
}

\abstract{
We quantize the open string in an arbitrary constant magnetic field with
a non factorized metric on a torus.
We then discuss carefully the vertexes which 
describe the emission of dipole open strings and closed strings 
in the non compact limit.
Finally we compute various stringy form factors which in the
compact case induces a K{\ae}hler and complex structure dependence and
suppression of some amplitudes with KK states.
}


\preprint{DFTTO-2010-8}

\begin{document}

\section{Introduction and conclusions}

The construction of ``phenomenological'' models of particle physics
based on D-branes embedded in supersymmetric string compactifications
has become a major direction in the development of String Theory 
(for review, see for instance 
\cite{Marchesano:2007de}).

In these D-brane scenarios non-perturbative corrections arise from D-instantons
and wrapped Euclidean branes. In principle all instantonic branes which can be
consistently included may contribute to the low energy effective action. If the
theory contains a gauge sector realized on D$(3+n)$-branes wrapped on a cycle
$\mathcal{C}$, then Euclidean branes E$(n-1)$ wrapped on $\mathcal{C}$
correspond to the instanton sectors of the gauge theory%
\footnote{The simplest case is represented by the D3/D(--1) system,
corresponding to $n=0$.} 
. 
Other Euclidean
branes, for instance those wrapped on a cycle $\mathcal{C}^\prime\not=
\mathcal{C}$, do not possess this interpretation and have been referred to as
``exotic'' or ``stringy'' instantons; they have been investigated over the last
 years in a rapidly growing literature (for a review, see for example
 \cite{Blumenhagen:2009qh}) .

This interest was sparkled by the realization that exotic instantons
might provide couplings which are forbidden in perturbation theory but
necessary for phenomenological applications; for instance, they have
been pointed out as possible sources of neutrino masses
or of certain Yukawa couplings in GUT models 
.  The relation of
``ordinary'' instantonic branes to the field-theoretical description
of instantons in supersymmetric gauge theories (as reviewed, for
instance, in \cite{Dorey:2002ik}) has been clarified in detail
\cite{Green:2000ke,Billo:2002hm}.
In the ordinary cases, the spectrum and interactions of the moduli, {\it i.e.} of the physical excitations of open strings with at least one end-point on the instantonic branes, reproduce the ADHM construction 
of the moduli space of gauge theory instantons.
In particular, let us consider the NS sector of the open strings with
one end on the D$(3+n)$ and the other end on the E$(n-1)$-brane. 
The world-sheet condition for physical states has the form%
\footnote{In order to be explicit, we consider a toroidal orbifold
  situation, and assume that the branes in the internal space are
  distinguished by their relative angles or magnetizations.}
\begin{equation}
 \label{vira}
L_0 - \frac 12 
= N_X + N_\psi +  \sum_{i=1}^3\frac{\nu_i}{2} = 0~,
\end{equation}
where $N_X$ and $N_\psi$ are the occupation numbers for the bosonic and
fermionic world-sheet oscillators, while the (positive) angles $2\pi\nu_i$
denote the twist eventually occurring in the three complex internal directions.
In writing (\ref{vira}) we have taken into account the $1/2$ contribution to the
zero-point energy from the four space-time directions, which are of
Neumann-Dirichlet type.
Ordinary E$(n-1)$-branes impose, in the internal space, the same boundary
conditions as the gauge D$(3+n)$-branes do. All twists $\nu^i$ therefore
vanish and the ground state $N_X=N_\psi=0$ is physical; being degenerate in the
non-compact directions, it corresponds to the moduli $w_{\dot\alpha}$ of the
ADHM construction. These bosonic mixed moduli enter in an essential way in the
ADHM constraints and, once these constraints are solved, they contain in
particular the size $\rho$ of the instanton solution.
The instanton profile and its dependence on ADHM parameters
 can be read from the amplitude \cite{Billo:2002hm}.
\begin{equation}
A_\mu(p;w,\bar w)=
\big\langle\,{\cal V}_{A_{\mu\nu}}\,\big\rangle_{\mbox{mixed disk}}
=
\langle V^{(-1)}_{\bar w}\,{\cal V}^{(0)}_{A^I_\mu}(-p)\,V^{(-1)}_{w}
\rangle
\label{instanton}
\end{equation}
exactly as the brane gravitational source profile can be read from the amplitude
(\cite{DiVecchia:1997pr})
\begin{equation}
h_{\mu \nu}(p)
=
\big\langle\,{\cal V}_{h_{\mu\nu}}\,\big\rangle_{\mbox{disk}} =
\langle h_{\mu\nu}(p)|{\rm D}p\rangle~~,
\label{graviton}
\end{equation}
where  $|{\rm D} p\rangle$ is the boundary state associated to the
given ${\rm D} p$ brane. 

The main result of this paper is the generating vertexes
(\ref{SDSzero}, \ref{SDSpi}, \ref{SDSclosed})
which allow to
compute all amplitudes involving two generic twisted states.
An immediate consequence shown in section \ref{sect:formfactors} is
that all three points mixed amplitudes
computed in literature have missed a momentum dependent
normalization factor\footnote{
Nevertheless similar factors have appeared in four points amplitudes,
see for example (\cite{Antoniadis:2000jv},\cite{Cvetic:2003ch}).}. 
This has a certain number of consequences.
\begin{itemize}
\item
The amplitude (\ref{instanton}) reads
\begin{align}
\label{instanton1} 
A^I_\mu(p;\bar w, w) = i\, (T^I)^{v}_{~u}\,p^\nu
\, \bar\eta^c_{\nu\mu}
\left(w_{\dot\alpha}^{~u}\,(\tau_c)^{\dot\alpha}_{~\dot\beta}\,
\bar w^{\dot \beta}_{~v}\right) 
\,e^{-i p\cdot x_0}
\,e^{-\oh R^2(\oh)\, \alpha' p_\nu p^\nu}
\end{align}
where the missed factor is the last one with $R^2(\oh)=4\ln 2$.
As it is clear from its dependence on $\alpha'$ it does not affect
much the infrared properties of the instantons and therefore in the
non compact limit the result found in literature such as for example 
(\cite{Billo:2007sw},\cite{Billo:2007py},
\cite{Billo':2010bd},\cite{Billo:2009di}) for the non compact cases are right. 
\item
Would it be only for the previous case it would appear that the
consequences are purely of principle, even if interesting.
There are nevertheless cases where the additional factor can have a
certain impact. As it is clear from the previous expression and from
eq. (\ref{D25DpGeneral}) the
additional factor is sensitive to the transverse momenta.
In the compact case  this means that the
amplitudes involving the creation of a couple of Kaluza-Klein
scalar particles are affected: these are the  amplitudes analogous   
to the fermionic ones computed in (\cite{Lust:2008qc}).
\item
Also the action of $D9/D5$ (\cite{Noi}) acquires higher order
corrections which can be useful to shed further light on the non
abelian Dirac-Born-Infeld action in the case of very strong magnetic fields.


\end{itemize}

The paper is organized as follows.
In section \ref{sect:openstring} we consider the quantization of open
string on a generic torus with a generic constant gravitational
background with a generic constant field strength in the Cartan subalgebra.
We do not however discuss the vacuum of the theory in the compact case
since it is not so straightforward and its use in the rest of the
paper  does not affect the main result in any way which cannot be
argued about.
In section \ref{sect:vertexes} we discuss the construction of the
vertexes in the old formalism where we describe the emission of dipole
strings from a single ``carrier'' dicharged one. We start from the
simplest case where the magnetic fields are only in two directions and
then we derive the generating function of all vertexes (Sciuto - Della
Selva - Saito vertex) for the general case. This vertex is a
generalization of the old result by Corrigan and Fairlie
(\cite{Corrigan:1975sn}).
We obtain also the corresponding 
emission vertex for closed string states. Using this result it would be
interesting to compute the boundary state with two twist fields
inserted using the technique developed in (\cite{Pesando:2009tt})
since boundary states have found many
applications such as and not only (\cite{Billo:1997eg},
\cite{DiVecchia:1997pr}, \cite{Billo:1998vr}, \cite{Billo:2002ff}, \cite{Bachas:2003sj}) (for a
review see for example \cite{Di Vecchia:1999rh}).

Finally  in section \ref{sect:formfactors} we use the formalism
developed to compute various three points 
mixed amplitudes which we interpret as form factors. 
We justify this interpretation and we discuss the difference between
amplitudes like $D25/D25'$ and $D25/D23$ as far as zero modes are
concerned.
Finally we compute the instanton form factor (\ref{instanton1}).

\section{Open string in arbitrary constant background}
\label{sect:openstring}
In this section we proceed to quantize the string on a generic torus 
with non factorized metric and generic constant magnetic
field. Nevertheless we do not care of describing the proper vacuum for
the zero modes in the compact case since it is not so straightforward
to build (\cite{Me}) and it is not needed for the rest of the paper.
Nevertheless a proper treatment of zero modes and Chan-Paton matrices
as in (\cite{Pesando:2009tt}) for the dipole strings is necessary to
solve some apparent inconsistencies (\cite{Di Vecchia:2006gg},
\cite{DiVecchia:2007dh}) and correctly explain some phenomena as the
gauge rank reduction in presence of discrete $B$
 (\cite{Pesando:2008xt}).

\subsection{The action}
The action for the spatial compact coordinates, in the following labeled by
$i,j=1, \dots,  {d}$, of  a bosonic string interacting with
constant gravitational and $2$-form backgrounds is given by:
\begin{eqnarray}
S_{bulk} \equiv - \int d\tau  \int_{0}^{\pi} d \sigma L  = - \frac{1}{4 \pi
  \alpha'}
\int d\tau \int_{0}^{\pi} d \sigma
\left[ G_{i j} \partial_{\alpha} X^i \partial_{\beta} X^j \eta^{\alpha
    \beta} - B_{i j} \epsilon^{\alpha \beta} \partial_{\alpha} X^i
\partial_{\beta} X^j \right]
\label{acti853}
\nonumber\\
\end{eqnarray}
where the world-sheet metric is $\eta_{\alpha \beta} =
\mbox{diag}(-1,1)$ and
$\epsilon^{01} =1$. We assume
$B$  and $G$  to be independent of $X$. 
Notice that in flat non compact space a
constant background for $B$ can be always gauged to  $B=0$ by
in the case of closed string but not
for open string
while $G$ can always be put in a diagonal form by a
change of coordinates but we will not do this since we are interested
in the compact case. 

We consider open strings and therefore add also an open string
background  consisting of a gauge fields describing 
an arbitrary field strength $\F_{i j}$ in the Cartan subalgebra.
This field strength can be diagonalized in color space.
Without a big loss of generality we can assume that
$\F_{i j}=F_{0\, i j} \uno_{N_0} \oplus F_{\pi\, i j} \uno_{N_\pi}$
where $\uno_N$ is the identity in a $N$ dimensional color subspace
 and we have not assumed
that $F_{0\,i j}$ and $F_{\pi\,i j}$ commute.
On this background we have different open string sectors: two dipole
string  and two dicharged ones. 
In particular of the two dicharged open strings we 
consider the string with the
constant commuting field strength $F_{0\,i j}$ on the $\sigma=0$
boundary and $F_{\pi\,i j}$ on the $\sigma=\pi$ ones \footnote{
In compact directions it is also sometimes necessary to add non
commuting Wilson lines, i.e. which do not belong to the Cartan subalgebra. 
These are included in the transition functions.}. 
The action of an open string in a
closed toroidal string background interacting with those gauge
fields is then given by
\begin{eqnarray}
S = S_{bulk} + S_{boundary}
\label{spluss}
\end{eqnarray}
where $S_{bulk}$ is given in Eq. (\ref{acti853}) and $S_{boundary}$ is
equal to:
\begin{eqnarray}
S_{boundary} 
&=&
 - q_0 \int d \tau A_{0\,i} \partial_{\tau} X^{i}
|_{\sigma =0} + q_{\pi} \int d \tau A_{\pi\,i} \partial_{\tau} X^{i}
|_{\sigma =\pi}
\end{eqnarray}
where $q_0$ and $q_{\pi}$ are the charges located at the two
end-points.

\subsection{Boundary conditions}
An open string in the previous metric background given by $E_{i j}=G_{i j}+B_{i j}$
and in presence of a constant background field 
$\hat F_{A\,i j}= 2\pi \alpha' q_A F_{A\,i j}$  $(A=0,\pi$) 
has  boundary conditions
\begin{equation}
G_{i j} X^{i} {}' + (B -\hF_{0})_{i j} \dot X^j |_{\sigma=0}=0
,~~~~
G_{i j} X^{i} {}' + (B -\hF_{\pi})_{i j} \dot X^j |_{\sigma=\pi}=0
\end{equation}
which can be written 
either as
\begin{equation}
\cE_{0\, j i} \partial_+ X^j -\cE_{0\, i j} \partial_- X^j|_{\sigma=0}=0
,~~~~
\cE_{\pi\, j i} \partial_+ X^j -\cE_{\pi\, i j} \partial_- X^j|_{\sigma=\pi}=0
\label{bc-open-noR}
\end{equation}
or 
as 
\begin{equation}
\partial_- X^i -R^i_{0  j} \partial_+ X^j|_{\sigma=0}=0
,~~~~
\partial_- X^i|_{\sigma=\pi} -R^ i_{\pi~ j} \partial_+ X^j =0
\label{bc-open}
\end{equation}
when we introduce
\begin{equation}
\cE_{0,\pi}=E^T +\hF_{0,\pi} = G +(\hF_{0,\pi} -B) = G+ \cF_{0,\pi}
\label{def-cE}
\end{equation}
and the orthogonal matrices
 \begin{equation}
R_0= \cE^{-1}_{0}  \cE^{T}_{0},~~~~
R_\pi= \cE^{-1}_{\pi}  \cE^{T}_{\pi},~~~~  
\end{equation}
 which satisfy the conditions
\begin{equation}
R_{0,\pi} G^{-1} R_{0,\pi}^T = G^{-1},~~~~
R_{0,\pi}^T G R_{0,\pi} = G
.
\end{equation}

\subsection{String quantization for a generic magnetic constant
  background on $T^{n}$.}
Given the boundary conditions (\ref{bc-open}) it is immediate to
write a solution of the equation of motion  
which respects the $\sigma=0$  ones as
\begin{equation}
X^i(\sigma,\tau)
=
x_0^i
+\oh \left( 
\Xln^i(\tau+\sigma)
+ (R_0)^i_j \Xrn^j(\tau-\sigma)
\right)
\end{equation}
The $\sigma=\pi$ boundary condition implies\footnote{
Notice that in $\Xln$ expansion there is also the momentum and that we
consider $x^i_0$ only as zero modes.
}
\begin{equation}
\Xln^{i '}(\tau+2\pi) = R^i_{~j}~\Xln^{j '}(\tau)
\end{equation}
with 
\begin{equation}
R=R_\pi^{-1} R_0
\Rightarrow
R G^{-1} R^T = G^{-1},~~~~
R^T G R = G
\label{R-ortho}
\end{equation}
To continue it is convenient to introduce the $R$ eigenvectors $v_a$
as
\begin{equation}
R^i_{~j} v_a^j = e^{-i 2\pi \nu_a} v_a^j,~~~~ a=\pm1, \dots \pm n
~~.
\label{R-eigen}
\end{equation}
where all eigenvectors (but one in odd dimensions) come in couples and
are labeled as
\begin{equation}
v_{-a}=v^a=v^*_a,~~~~
\nu_{-a}=-\nu_a
\end{equation}
with $a>0 \rightarrow \nu_a\ge 0$ 
\footnote{If the dimension of the space is odd we take $\nu_{a=}=0$ 
and $v_{a=0}$ to be real.}
\footnote{We use the convention that while indexes $a,b$ label any
  eigenvalues  $c,d$ label those different from zero, i.e. $\nu_c,\nu_d\ne0$
  while $e,f,g$ label those equal to zero, i.e.
  $\nu_e,\nu_f,\nu_g=0$} 
and are normalized as
\begin{equation}
v_{a}^\dagger G v_b =v_{-a}^T G v_b = \delta_{a,b}
~~.
\end{equation}

For future convenience we write also the spectral decomposition of the
$R$ and $G^{-1}$ matrices as
\begin{equation}
(G^{-1})^{i j}= \sum_a v_a^i~v_{-a}^j
,~~~~
(R G^{-1})^{i j}= \sum_a v_a^i~e^{-i 2\pi \nu_a}~v_{-a}^j
\label{spectralGR}
\end{equation}

It is then possible to write the expansions for $\Xln$ and $\Xrn$ as
\begin{eqnarray}
\Xln^i(\xi)
&=&
\sum_a v^i_a \sum_n 
~\beta_{n}^a ~e^{-i (n+\nu_a)\xi}
\nonumber\\
\Xrn^i(\xi)
&=&
\sum_a (R_0 v_a)^i \sum_n 
~\beta_{n}^a~ e^{-i (n+\nu_a)\xi}
\end{eqnarray}
with the understanding that when $n+\nu_a=0$ then $e^{\pm i (n+\nu_a)\xi}=\xi$.
We can therefore write the complete expansion of the string field in a
way similar to the usual one as
\begin{align}
X^i(\sigma,\tau)
&=
x_0^i
+i \sqrt{2\alpha'} 
\sum_{a,n / n+\nu_a\ne 0} \frac{\alpha_n^a}{n+\nu_a}  ~\Psi^i_{n\, a}(\sigma,\tau)
+ \sqrt{2\alpha'}  \sum_{e / \nu_e=0} \alpha_0^e ~\Psi^i_{0\, e}(\sigma,\tau)
\end{align}
where the we have defined the eigenfunctions
\footnote{
For the zero modes we could also use a more symmetric formulation with
respect to exchanges  
$\sigma=0 \leftrightarrow \sigma=\pi$ and
$\hF_0 \leftrightarrow -\hF_\pi$ using
$
\hat \Psi_{0\, e}^i(\sigma,\tau)
=
\oh
\left(
v_e^i~(\tau+\sigma-\frac{\pi}{2})
+
(R_0 v_e)^i~ (\tau-\sigma-\frac{\pi}{2})
\right)
$
but the computation are less straightforward and the result as complex.
}
\begin{align}
\Psi_{n\, a}^i(\sigma,\tau)
&=
\oh
e^{-i (n+\nu_a)\tau}
\left(
v_a^i~ e^{-i (n+\nu_a)\sigma}
+
(R_0 v_a)^i~ e^{+i (n+\nu_a)\sigma}
\right)
,~~~~
n+\nu_a \ne 0
\nonumber\\
\Psi_{0\, e}^i(\sigma,\tau)
&=
\oh
\left(
v_e^i~(\tau+\sigma)
+
(R_0 v_e)^i~ (\tau-\sigma)
\right)
,~~~~
\nu_e=0
\label{Eigenfunctions}
\end{align}

\subsubsection{Completeness of the $\Psi_{n\, a}$}
We want now show that $\Psi_{n\, a}^i(\sigma,\tau)$ form a complete
basis for all functions with the same boundary conditions.
To this purpose we define the product\footnote{
We take $\int_0 d\sigma~ \delta(\sigma)=1$ since taking
$\int_0 d\sigma~ \delta(\sigma)=\oh$  as in \cite{PRD5249}
would not reproduce the dipole case. Similarly for $\sigma=\pi$.
\COMMENTOO{\bf meglio}

} 
\begin{equation}
\langle X^i(\sigma,\tau), Y^j(\sigma,\tau)\rangle
=
\int_0^\pi d \sigma~
X^{i *}(\sigma,\tau) 
~\left[
G ~ \left(
\stackrel{\rightarrow}{\partial}_\tau  -
\stackrel{\leftarrow}{\partial}_\tau  
\right)
+\cF_0 ~\delta(\sigma)
-\cF_\pi~\delta(\sigma-\pi)
\right]_{i j} 
~Y^j(\sigma,\tau)
\label{AntiHermProd}
\end{equation}
It is worth noticing that this product shall also be used for the
subspaces where $\cF_0=\cF_\pi$ and  are not factorisable because of
the metric.

Let us now consider a generic field
\begin{equation}
\Phi^i(\sigma,\tau)
=
\phi^i
+ \sum_{a,n} \phi^a_n   ~\Psi^i_{n\, a}(\sigma,\tau)
\end{equation}
and try to extract the coefficients $\phi$. We get immediately that
\begin{align}
\langle \Psi_{a\,n} , \Phi \rangle
&=
-i \pi (n+\nu_a)~\phi^a_n
&
n+\nu_a\ne 0
\nonumber\\
\langle \cE^{-1}_\pi G v_e , \Phi \rangle
&=
\pi \phi^e_0
& 
\nu_e=0
\label{defI0}
\end{align}
from which we can easily extract the coefficients $\phi^a_n$. 
The analogous expressions for  $\phi^i$ give
\begin{align}
\langle \Psi_{0\,e} , \Phi \rangle
&=
-\pi~ v_e^\dagger \left( \cE_\pi~\phi 
                     + \pi ~\cF_\pi~v_f~ \phi^f_0\right)
& 
\nu_e=0
\nonumber\\
\langle \cE^{-1}_\pi ~G ~v_c , \Phi \rangle
&=
-v_c^\dagger~G~\cE_\pi^{-T}~\Delta\cF~\phi
&~ 
\label{defI}
\end{align}
with $\Delta \cF= \cF_\pi-\cF_0$.
From  the previous equations it
is possible in principle to recover the $\phi^i$ coefficients 
since the equations in the last two lines are
as many as the components of $\phi^i$ and are independent 
and are therefore sufficient but the result is rather not instructive.
In order to compute the commutation relation we resort therefore to
the trick of computing the commutation relations among the previous
quantities (\ref{defI0}, \ref{defI}) when we take $\Phi=X$ 
and then extract those among the $x^i_0$.

\subsubsection{Commutation relations}
Using the canonical commutation relation
\begin{equation}
[X^i(\sigma,\tau)~,~P_j(\sigma',\tau)]
= i \delta^i_j~ \delta(\sigma-\sigma')
\label{CanComRel}
\end{equation}
we can deduce the commutation relations (for details see appendix \ref{appDetailsQuant})
\begin{align}
[\alpha^a_n,~\alpha^b_m]=& \delta_{a+b,0}~ \delta_{m+n,0}~(n+\nu_a)
~~~~~n+\nu_a,m+\nu_n\ne 0
\nonumber\\
[\alpha_{0}^e,~\alpha^a_n]=&0
~~~~~n+\nu_a\ne 0
\nonumber\\
[x^i,~\alpha^c_n]=&0
\nonumber\\
[\alpha^e,~\alpha^f_0]=&0
\nonumber\\
[\alpha^e_0,~x^i_0]
=& - i ~\sqrt{2\alpha'}~(\cE_\pi^{-1} G v_{-e})^i
\nonumber\\
[x^i_0,x^j_0]=&
+i ~2\pi\alpha'~ \sum_{c / f_c\ne 0} w^i_c \frac{1}{f_c} ~(w^\dagger_c)^j
\nonumber\\
&-i ~2\pi\alpha'~ \sum_{c,f / f_c\ne 0,\nu_f=0}
\left[ w^i_c \frac{1}{f_c} ~(w^\dagger_c ~\cE_\pi^T ~v_f) 
(v^\dagger_f~G~\cE_\pi^{-T})^j
+
(\cE_\pi^{-1} G v_f)^i (v_f^\dagger~\cE_\pi~w_c)
\frac{1}{f_c} ~(w^\dagger_c)^j
\right]
\nonumber\\
&+i ~2\pi\alpha'~ 
\sum_{f,g / \nu_f=\nu_g=0}
(\cE_\pi^{-1} G v_f)^i
\left[
v^\dagger_f~\cF_\pi~v_g 
+
\sum_{c/ f_c\ne 0}
(v_f^\dagger~\cE_\pi~w_c)\frac{1}{f_c} ~(w^\dagger_c ~\cE_\pi^T ~v_g) 
\right]
(v^\dagger_g~G~\cE_\pi^{-T})^j
\label{CommRel}
\end{align}
The last commutation relations can also be rewritten in a more compact
form by using the spectral decomposition of $G^{-1}$ as
\begin{align}
[x^i_0,x^j_0]=&
+i ~2\pi\alpha'~ 
\sum_{c,d / \nu_c,\nu_d\ne 0}
(\cE_\pi^{-1} G v_c)^i
\left[
\sum_{c_1/ f_{c_1}\ne 0}
(v_c^\dagger~\cE_\pi~w_{c_1})\frac{1}{f_{c_1}} ~(w^\dagger_{c_1} ~\cE_\pi^T ~v_d) 
\right]
(v^\dagger_d~G~\cE_\pi^{-T})^j
\nonumber\\
&+i ~2\pi\alpha'~ 
\sum_{f,g / \nu_f=\nu_g=0}
(\cE_\pi^{-1} G v_f)^i
~v^\dagger_f~\cF_\pi~v_g 
~(v^\dagger_g~G~\cE_\pi^{-T})^j
\label{CommRel0}
\end{align}
If we introduce the decomposition\footnote{
Notice however that the same decomposition does not simplify the non
zero modes commutation relations as better detailed in footnote \ref{foot:diff_proj}.
}
\begin{equation}
x_0^i= \sum_a (\cE_\pi^{-1}G v_a)^i\, x^a_0
~~\Leftrightarrow~~
x^a_0= v_a^\dagger \cE_\pi x_0
\end{equation}
then all the previous non vanishing commutation relations involving
$x_0$ boil down to
\begin{align}
[x^e_0\,,\,\alpha^{-f}_0]&= i \sqrt{  2\alpha'}\, \delta_{e,f}
\nonumber\\
[x^e\,,\,x^{-f}_0]&= i 2\pi\alpha'\, v_e^\dagger \cF_\pi v_{f}
\nonumber\\
[x^c\,,\,x^{-d}_0]&= i 2\pi\alpha'\,
\sum_{c_1/ f_{c_1}\ne 0}
(v_c^\dagger~\cE_\pi~w_{c_1})\frac{1}{f_{c_1}} ~(w^\dagger_{c_1} ~\cE_\pi^T ~v_d)
\end{align}

In the previous commutation relations and in (\ref{CommRel},
\ref{CommRel0}) 
new vectors $\{w_c\}$ appear. They are related the eigenvectors of $\Delta \cF$ and are an
expression of the fact that $x_0$s feel mostly $\Delta \cF$ while the
other non zero modes  feel $\Delta R=R_\pi-R_0$.
We define therefore the $\Delta \cF$ eigenvectors
\begin{equation}
(\Delta\cF)_{i j}~ w^j_a= f_a~ G_{i j}~w^j_a
\end{equation}
as for $R$, all eigenvectors (but one in odd dimensions) come in
couples and are labeled as 
\begin{equation}
w_{-a}=w^*_a,~~~~
f_{-a}=f_a^*=-f_a
\end{equation}
and normalized as
\begin{equation}
w_{a}^\dagger G w_b =w_{-a}^T G w_b = \delta_{a,b}
\end{equation}
It is therefore possible to write the spectral decomposition of the
$\Delta\cF$ and $G^{-1}$ matrices as
\begin{equation} 
(G^{-1})^{i j}= \sum_a w_a^i~w_{-a}^j
,~~~~
\Delta\cF_{i j}= \sum_a (G w_a)_i~f_a~(G w_{-a})_j
\label{spectralGF}
\end{equation}

It is then possible to check that 
$\cE_0^{-1} G ~ker(\Delta R)=\cE_\pi^{-1} G ~ker(\Delta R)
=ker(\Delta\cF)$.
This happens since 
\begin{equation}
R_\pi-R_0
=
2 \cE_0^{-1} ( \cF_\pi -\cF_0) \cE_\pi^{-1} G
=
2 \cE_\pi^{-1} ( \cF_\pi -\cF_0) \cE_0^{-1} G
\end{equation}
therefore for any $v_f$ such that $\nu_f=0$, i.e. $ R_0 v_f=
R_\pi v_f$, the vector $\cE_\pi^{-1} G v_f$ belongs to $ker(\Delta\cF)$
and similarly for $G^{-1} \cE_\pi w_f$ with $f_f=0$.
We can therefore use the $\cE_\pi^{-1} G v_f$ as basis for 
$ker(\Delta\cF)$ even if they are not orthogonal.
This explains why they appear  in the $x_0$ commutation relations. 
Notice finally that generically it is not true that
$\cE_0^{-1} G ~[ ker(\Delta R) ]^\perp = [ker(\Delta\cF)]^\perp$ even if they have the
same dimension.

Finally we can check what happens of the $x_0$ commutation relations in
special cases.
When $ker(\Delta\cF)=\emptyset$, i.e.
there are no subspaces where $\cF_0=\cF_\pi$
the previous commutation relation for the $x_0$ reduces to
\begin{eqnarray}
{}[x^i_0, x^j_0] 
&=&
i 2\pi\alpha' [( \hF_{\pi}-  \hF_{0})^{-1}]^{i j}
\end{eqnarray}
while in the case where $[ ker(\Delta\cF)] ^\perp=\emptyset$, i.e.
$\cF_0=\cF_\pi$  it reduces to
\begin{eqnarray}
{}[x^i_0, x^j_0] 
&=&
i 2\pi\alpha' [\cE_\pi^{-1} \cF_\pi \cE_\pi^{-T}]^{i j}
=
i 2\pi \alpha' \Theta^{i j}
\end{eqnarray}
upon the use of the spectral decomposition of $G^{-1}$ given in eq.
(\ref{spectralGR}).
Under the same assumption comparing with the usual dipole string expansion
we see that $ \sqrt{2 \alpha'} p^i= (\cE_\pi^{-1} G ~v_e)^i~ \alpha^e_0$
so that the commutation $[\alpha^e_0,~x^i]$ becomes
$[p^j,~x^i]= -i \cG_\pi^{j i}= -i \cG_0^{j i}$ as expected.

\subsubsection{The vacuum and the spectrum}
The vacuum of the non zero modes (here we consider only $x^i_0$ as
zero modes) can be defined as
\begin{equation}
\alpha^a_n |0_e,\, 0_{\hat c},\, 0_\alpha\rangle 
~= 
\alpha^e_0 |0_e,\, 0_{\hat c},\,0_\alpha\rangle 
~=0
~~~~
n+\nu_a>0
\end{equation}
Instead for the $x^i_0$ we must find a maximum commuting set among
themselves and also with $\alpha^e_0$. The $x_0$ commuting with
$\alpha^e_0$ are the combinations
$x^c_0=v_c^\dagger~\cE_\pi~ x_0$ which nevertheless 
do not commute among themselves.
We can form the proper number of commuting linear combinations of
$x^c_0$, let us call them $\hat \pi_{\hat c\,0}$ ($ {\hat c}=1,\dots ,\oh dim~[ker(\Delta\cF)]^\perp$) by diagonalizing at
$2\times 2$ blocks the previous antisymmetric matrix. At the same time 
we get the corresponding ``coordinates'' $\hat \xi^{\hat c}_ 0 $ with the
usual algebra $[\hat \xi^{\hat c}_0,\hat \pi_{\hat d\, 0 }]= i
\delta^{\hat c}_{\hat d} $.
Finally we can define the vacuum for the zero modes by requiring that
\begin{equation}
\hat \pi_{\hat c\,0} |0_e,\, 0_{\hat c},\,0_\nu\rangle=0
~~~~ {\hat c}=1,\dots ,\oh \mbox{dim}~[ker(\Delta\cF)]^\perp
\end{equation}
This definition of the $x_0$ vacuum is the right one only for the non
compact case since on the torus $x_0$s are not anymore good operators
and the vacuum definition is more complex (\cite{Me}).
The spectrum is then given by
\begin{align}
\prod_a \prod_{n=1}^\infty ( \alpha_{-n}^a)^{N_{n\,a}} 
|k_e, \kappa_{\hat  c}, 0_\alpha\rangle 
\label{spectrum}
\end{align}
with normalization
\begin{equation}
\langle k'_e,\kappa'_{\hat  c}, 0_\alpha |
k_e, \kappa_{\hat  c}, 0_\alpha \rangle
= (2\pi)^d\, \delta^{\mbox{dim}\,ker\,(\Delta\cF)}(k'_e-k_e)
\, \delta^{\mbox{dim}\,(ker\,(\Delta\cF))^\perp}
(\kappa'_{\hat  c}-\kappa_{\hat  c})
\end{equation}
It is worth stressing that also in half the ``twisted''
directions there is a ``momentum'' $\kappa_{\hat  c}$ since in the non compact
case there is still a translational invariance roughly because it is
possible to choose a gauge where $\cF_0 -\cF_\pi$ is derived from a
potential depending on only half the number of coordinates.
While this choice is still possible in the compact case the transition functions
associated to the gauge bundle break this remaining translational invariance
and we are left with a finite Landau levels degeneracy. 
Notice however that this  $\kappa_{\hat  c}$ is not a true momentum since it
does not appear in the Hamiltonian and it is actually a label for the
degeneration. Another way of saying this is that the choice of using
$\hat \pi_{\hat c\,0}$ as annihilators is completely arbitrary and we
could and well have chosen a mixture of $\hat \pi_{\hat c\,0}$ and
$\hat \xi^{\hat c}_ 0 $ or any other possible combination of
commuting operators.

\subsubsection{Energy momentum tensor}
We can now compute the energy momentum tensor to be
\footnote{
Remember that the extra term is due to the presence of a subleading
singularity in the double derivative of the propagators. 
} 
\begin{align}
T&= 
-\frac{1}{\alpha'} \partial_+ X^T G \partial_+ X
=
-\frac{1}{4\alpha'} \partial_+ X_L^T G \partial_+ X_L
=
\sum_n L_n e^{-i n (\tau+\sigma) }
\nonumber\\
L_n &= \sum_a \oh \left[ \sum_k : \alpha^a_{n-k} \alpha^{-a}_k: 
+\delta_{n,0}  \nu_a(1-\nu_a)
\right]
\label{dichargedT}
\end{align}
It is interesting to notice that the shift can be understood as due to
representing the dipole energy momentum tensor in the dicharged string
as explained in section \ref{ExamplesV}.

\subsubsection{OPEs}
The basic OPE for the non zero modes  is given by ($z= e^{\tau_E+i \sigma}$)
\begin{eqnarray}
\Xln^{i(+)}(z)~\Xln^{j(-)}(w)
&=&
-2\alpha' \sum_a v^i_a~ \hat g_{-\nu_a}\left(\frac{w}{z}\right)~ v^j_{-a}
 ~~~~
|z|>|w|
\end{eqnarray}
where we have defined the ``propagator'' as
\begin{equation}
 \hat g_{\nu_a}\left(z\right)
=-\sum_{n-\nu_a>0} \frac{1}{n-\nu_a} z^{n-\nu_a}
~~~~
|z|<1~,~
-\pi<arg(z)<\pi
\label{propagator}
\end{equation}
From the previous basic OPE we can derive all the others such as for
example\footnote{
\label{foot:diff_proj}
These OPEs and the zero modes commutation relations show why it is
difficult to define in a unique way well adapted coordinates: in fact
the operators with the simplest relations are 
$x^a_0= v_a^\dagger \cE_\pi x_0$ , $\Xln^a(z)=  v_a^\dagger G \Xln(z)$
and
$\Xrn^a(\bar z)=  v_a^\dagger G\cE_0^{-T} \cE_0 \Xrn(\bar z)$.
}
\begin{eqnarray}
\Xln^{i(+)}(z)~\Xrn^{j(-)}(\bar w)
&=&
-2\alpha' \sum_a v^i_a~ \hat g_{-\nu_a}\left(\frac{\bar w}{z}\right)~ (R_0 v_{-a})^j
\end{eqnarray}

\subsubsection{$\Omega$ operator}
It is also possible to define the $\Omega$ operator which exchanges
the string boundaries and therefore maps a string $X$ with boundary
field strength $(\cF_0,\cF_\pi)$  into a (generically different) string
$\tilde X$ with boundary field strength $(\tilde \cF_0=
-\cF_\pi,\tilde \cF_\pi=-\cF_0)$. The minus sign is due to the
orientation of the boundaries.
Starting from the expected  relation
\begin{align}
\Omega\, X(\sigma,\tau) \, \Omega^{-1}
=
\tilde X(\pi-\sigma,\tau)
\end{align}
and the consequences of the relation between $\tilde \cF$ and $\cF$ 
\begin{align}
\tilde R_0 = R_\pi ^{-1}
&,
\tilde R_\pi = R_0 ^{-1}
\nonumber\\
\tilde v_a &= R_0\, v_a e^{i\pi \nu_a}
\end{align}
(no sum over $a$ in the last equation)
we find that the $\Omega$ action on the operators is given by
\begin{align}
\Omega\, \alpha^a_n \, \Omega^{-1} &= (-)^n \tilde  \alpha^a_n
\nonumber\\
\Omega\, x_0 \, \Omega^{-1} &=  \tilde  x_0 + \sqrt{2\alpha'} \pi
\frac{1-\tilde R_0}{2} \tilde v_f \, \tilde \alpha^f_0
\end{align}

\section{Vertexes and the SDS vertex
}
\label{sect:vertexes}
In order to explain the derivation in a simple setting we start
considering the tachyonic vertexes on $\R^2$ with coordinates $(x^1,x^2)$.
We consider only two magnetized branes with different magnetization,
i.e. 
$\ff_0=(\hF_0-B)_{1 2}$  and $\ff_\pi=(\hF_\pi-B)_{1 2}$ 
so the gauge group is $U(2)$ and the dicharged string describes 
the matter in the bifundamental.

In this formalism (\cite{Ademollo:1974fc}) 
we have a carrier string which is the dicharged
string coupled to the two different magnetic fields. 
From this dicharged string we then
describe the emission of the dipole strings from the two boundaries.
Exactly as in the old superstring formalism 
where it was natural to describe the emission of NS states from a
R string.
Differently from the emission of NS states from a R string we have
here two non equivalent boundaries and therefore we can emit states of two
different dipole strings.

To give the complete picture is also important to describe the
coupling of the dicharged string with the closed string: it turns out
that it is exactly this coupling which constrains how the splitting in
left and right zero modes happens.

We start writing the tachyonic vertexes explicitly, then we give the
arguments that lead to their expressions 
and we construct the generator of all vertexes, i.e. the
Sciuto-Della Selva-Saito vertex.
Finally we discuss all the consistency requests we expect these
vertexes must satisfy.

\subsection{The tachyonic vertexes for $U(1)\times U(1)$ on $\R^2$}
\label{sect:tach-vert-R2}
In the following we work with flat complex coordinates but we do not
denote this explicitly while
in appendix \ref{app:Dipoleopen} 
 we are more pedantic and denote the flat vectors explicitly by an underline:
for examples the magnetic field in flat coordinates
$\underline{\ff}_0=(\hF_0-B)_{1 2}/ \sqrt{\det G}$ wil be written as $\ff_0$.
The part of  the tachyonic vertexes associated with the two twisted directions 
for the emission of a tachyonic dipole string
excitation from the $\sigma=0$, $\sigma=\pi$ boundary and for the
closed string tachyon read
\begin{align}
V_{T_0}(x, k)  
\propto & 
e^{-\ce \Delta_0(k)}
x^{-\Delta_0(k)}
:e^{i ( \bar k Z(x,x)+ k \bar Z(x,x) )}:
\label{VT0}
\\
V_{T_\pi}(y, k)  
\propto&  
e^{-\ce \Delta_\pi(k)}
|y|^{-\Delta_\pi(k)}
:e^{i ( \bar k Z(y,\bar y)+ k \bar Z(y, \bar y) )}:
\label{VTpi}
\\
W_{T_c}(z,\bar z, k_L,k_R)  
\propto&  
e^{-\ce \Delta_c(k_L)}
z^{-\Delta_c(k_L)}
:e^{i ( \bar k_L Z_L(z)+ k_L \bar Z_L(z) )}:
\nonumber\\
&
e^{-\ce \Delta_c(k_R)}
{\bar z}^{-\Delta_c(k_R)}
:e^{i ( \bar k_R Z_R(\bar z)+ k \bar Z_R(\bar z) )}:
\label{WTc}
\end{align}
where we have introduced the (partial) conformal dimensions
\begin{equation}
\Delta_0(k)=2\alpha' \cos^2 \gamma_0 ~k \bar k,
~~~~
\Delta_\pi(k)=2\alpha' \cos^2 \gamma_\pi ~k \bar k,
~~~~
\Delta_c(k)=2\alpha' ~k \bar k
\end{equation}
with the complex momentum given by $k=(k_1+i k_2)/\sqrt{2}$ 
and with 
$e^{i \gamma_{0,\pi}}= (1+i \ff_{0,\pi}) /\sqrt{1+\ff_{0,\pi}^2}$. 
We have also defined
the positive constant (for $0<\nu<1$  the positiveness follows from the
$\psi$ representation given in eq.(\ref{psi_sum}))
\begin{equation}
\ce= -\frac{1}{2}(\psi(1-\nu)+\psi(\nu)- 2 \psi(1))
\end{equation}
where $\nu= |\gamma_0-\gamma_\pi|/\pi$ and  $\psi(z)= \frac{d \ln \Gamma(z)}{d z}$ is the digamma function.
The normal ordering is then better defined splitting zero and non zero
modes and it amounts to
\begin{align}
&:e^{i ( \bar k Z(x,x)+ k \bar Z(x,x) )}:
=
~:e^{i \cos \gamma_0 ( \bar k e^{i \gamma_0} \Zln(x)
+ k e^{-i \gamma_0}  \bZln(x) )}:
~e^{i ( \bar k z_0+ k \bar z_0 )}
\nonumber\\
&:e^{i ( \bar k Z(y,y)+ k \bar Z(y,y) )}:
=
~:e^{i \cos \gamma_\pi ( \bar k e^{i \gamma_\pi} \Zln(y)
+ k e^{-i \gamma_\pi} \bZln (y) )}:
~e^{i ( \bar k z_0+ k \bar z_0 )}
\end{align}
for open string vertexes and to
\begin{eqnarray}
&&:e^{i ( \bar k_L Z_L(z)+ k_L \bar Z_L(z) )}:
= 
~:e^{i ( \bar k_L Z_{L \nzm}(z)+ k_L \bar Z_{L \nzm}(z) )}:
~e^{i ( \bar k_L \zlz+ k_L \bzlz )}
\nonumber\\
&&:e^{i ( \bar k_R Z_L(z)+ k_R \bar Z_L(z) )}:
= 
~:e^{i ( \bar k_R Z_{L \nzm}(z)+ k_R \bar Z_{L \nzm}(z) )}:
~e^{i ( \bar k_R \zrz+ k_R \bzrz)}
\end{eqnarray}
for the closed string one.
In  the previous definition of the normal ordering for the closed
string vertex we introduced the left and right zero modes $\zlz,
\zrz$.
The computation given in the next subsection is independent on this
splitting since in the $Z, \bar Z$ expansion 
there are no momentum operators which could be sensitive to this
splitting as it happens also whenever $ker \Delta\cF = \emptyset$.
The splitting can nevertheless determined by requiring that two closed
string vertexes (\ref{WTc}) have the same OPE as the usual ones which is one of
the consistency requirements we write down in section
\ref{sect:consistency} and we verify in appendix \ref{AppDetailsVertexes}.
Therefore here we can simply state the result 
\begin{equation}
\zlz=(1+i \ff_\pi) z_0,~~~~
\zrz=(1-i \ff_\pi) z_0.
\label{z0l-z0r-noncomp}
\end{equation}

\subsubsection{Heuristic derivation of the vertexes.}
To derive heuristically the previous vertexes 
we use the same approach used in (\cite{Hamidi:1986vh}).
The idea is to define the vertex for the emission of a dipole
state from the dicharged string starting from a regularized version of
the naive vertex, then eliminate the divergence by dividing 
it by the normalization required by the
emission vertex of the same state from the corresponding dipole
string 
and finally let the cutoff used to regularize the vertex to zero.

We start by writing the regularized version of the naive emission
vertex for the dipole tachyon from the $\sigma=0$ boundary of the
dicharged string.
The dipole tachyon is emitted from the dicharged string and
the not normal-ordered but point splitted emission operator can be
written as
\begin{align}
[ V_{T_0}(x, k) ]_{p.s.}  
= & 
e^{i [ \bar k\, ( Z^{(-)} (x e^{-\eta},x e^{-\eta}) + Z^{(+)}(x,x) )
+ k\, (\bar Z^{(-)}(x e^{-\eta},x e^{-\eta}) + \bar Z^{(+)}(x,x) ) ]}
\end{align}
Using the boundary conditions this expression can be rewritten as 
\begin{align}
= &
~\exp\{i \cos \gamma_0 [ \bar k e^{i \gamma_0} 
               ( \Zln^{(-)}(x e^{-\eta}) +    \Zln^{(+)}(x) )
\nonumber\\
&~~~~~~
+ k e^{-i   \gamma_0} 
               ( \bZln^{(-)}(x e^{-\eta}) +   \bZln^{(+)}(x) )
]\}
~e^{i ( \bar k z_0+ k \bar z_0 )}
\label{VT0reg}
\end{align}
where $\eta>0$ and, as usual, $\Zln^{(-)}$ is the part of the field $\Zln$
containing the creator operators  with the exclusion of the zero modes
$z_0, \bar z_0$.
It is then immediate to rewrite the previous expression as a
normal-ordered one 
\begin{align}
[ V_{T_0}(x, k)  ]_{p.s.}
= &
\exp\left \{ 
\cos^2\gamma_0~ \alpha'~k \bar k 
\left ( \hat g_\epsilon(e^{-\eta})+ \hat g_{-\epsilon}(e^{-\eta}) \right)
\right\}
\nonumber\\
&~:\exp\Big\{
i \cos \gamma_0 \big [ \bar k e^{i \gamma_0} 
               ( \Zln^{(-)}(x e^{-\eta}) +    \Zln^{(+)}(x) )
\nonumber\\
&~~~~~~
+ k e^{-i   \gamma_0} 
               ( \bZln^{(-)}(x e^{-\eta}) +    \bZln^{(+)}(x) )
\big ] \Big \} :
\nonumber\\
&
~e^{i ( \bar k z_0+ k \bar z_0 )}
\label{VT0regnorm}
\end{align}
where we have used the ``propagator'' $\hat g_\epsilon(z)$ defined in
eq. (\ref{propagator}).

We want now to eliminate the divergence which appears as $\eta
\rightarrow 0^+$ and we want to do this
by comparing with the analogous procedure we can follow for the usual
emission vertex from the dipole ``carrier'' string in order to
subtract as little as possible.
Let us therefore perform the same operations as above 
on the usual vertex operator which
describes the emission of the same dipole tachyon from
the corresponding dipole string. 
This dipole string is described by $Z^{(0)}$ and
$\bar Z^{(0)}$ and has the tachyon itself we are emitting among the
excitations. 
If we start from the analogous expression of (\ref{VT0reg}) 
where we use $Z^{(0)}$ and $\bar Z^{(0)}$ in place of $Z$ and $\bar Z$ we find
\begin{align}
[ V_{T_0}^{(0)}(x, k)  ]_{p.s.}
= &
\exp\left \{ 
\cos^2\gamma_0~ \alpha'~k \bar k 
\left ( 2 \hat g_0(e^{-\eta})+2 \ln x  \right)
\right\}
\nonumber\\
&~:\exp\Big\{
i \cos \gamma_0 \big [ \bar k e^{i \gamma_0} 
               ( \Zln^{(0,-)}(x e^{-\eta}) +    \Zln^{(0,+)}(x) )
\nonumber\\
&~~~~~~
+ k e^{-i   \gamma_0} 
               ( \bZln^{(0,-)}(x e^{-\eta}) +    \bZln^{(0,+)}(x) )
\big ] \Big \} :
\nonumber\\
&
~e^{i ( \bar k z_0^{(0)}+ k \bar z_0^{(0)} )}
~e^{ 2\alpha' ( \bar k p^{(0)} + k \bar p^{(0)} ) \ln x}
.
\label{V0T0reg}
\end{align}
Dividing this point splitted expression for
the non-operatorial factor in the first line 
\begin{equation}
\cN_0(\eta)
=
\exp\left \{ 
\cos^2\gamma_0~ \alpha'~k \bar k 
\left ( 2 \hat g_0(e^{-\eta})+2 \ln x  \right)
\right\}
\end{equation}
we get a regularized vertex
and letting $\eta\rightarrow 0^+$
 we recover the usual vertex operator for the emission from the
 dipole string
\begin{equation}
 V_{T_0}^{(0)}(x, k)  
=
\lim_{\eta \rightarrow 0^+} \cN_0^{-1}(\eta)
[ V_{T_0}^{(0)}(x, k)  ]_{p.s.}
\end{equation}
It is worth stressing that the zero modes 
$z_0^{(0)}$ and $\bar z_0^{(0)}$ do not give any contribution even if
 they do commute between themselves because they are not split into
 different exponentials and that the factor $\cos^2 \gamma_0$ comes
 from rewriting the vertex using the left moving part only.

If we divide the vertex operator (\ref{VT0regnorm}) for the same
factor written in the first line of eq. (\ref{V0T0reg}) we
can still take the $\eta\rightarrow 0^+$ since the two dimensional UV
divergences are the same and
the result is exactly the vertex given in eq. (\ref{VT0})
\begin{equation}
 V_{T_0}(x, k)  
=
\lim_{\eta \rightarrow 0^+} \cN_0^{-1}(\eta)
[ V_{T_0}(x, k)  ]_{p.s.}
\end{equation}

For the emission vertex from the $\sigma=\pi$ boundary the same
procedure works:
the regularized normal-ordered vertex written in term of the untwisted
fields $Z^{(\pi)}(y,y)$ and $\bar Z^{(\pi)}(y,y)$ must be divided by 
the product of the non-operatorial factor
$e^{ 
\cos^2\gamma_0~ \alpha'~k \bar k 
\left ( 2 g_0(e^{-\eta})+2 \ln |y|  \right)
}$
in order to give the usual emission vertex. 
Another way of getting this result is to use the twist operator $\Omega$.

For the closed string we proceed again as before and we compare with
the usual emission vertex expressed through the closed string fields
$Z^{(c)}$ and $\bar Z^{(c)}$ where the zero modes are treated as they
were independent. The normalization factor is easily computed to be
\begin{align}
\exp\left \{ 
 \alpha'~k_L \bar k_L 
\left ( 2 \hat g_0(e^{-\eta})+2 \ln z  \right)
\right\}
\exp\left \{ 
 \alpha'~k_R \bar k_R 
\left ( 2 \hat g_0(e^{-\eta})+2 \ln \bar z  \right)
\right\}
\end{align}

We are however left with the problem of how to split the zero modes
$z_0$ and $\bar z_0$ into left and right zero modes $z_{0 L}$, $z_{0  R}$.
As stated before and verified in appendix \ref{AppDetailsVertexes}
the splitting can determined by requiring that two closed
string vertexes (\ref{WTc}) have the same OPE as the usual ones.

\subsection{The SDS vertex and the generic vertex.}
We want now to reproduce the steps done in the previous section 
for all vertexes at once: this can be done using a generator
functional of the vertexes.

Following the spirit of the Sciuto-Della Selva-Saito approach we
define a generator functional for the vertexes describing the
 emission of dipole states from the $\sigma=0$ boundary of  the 
dicharged string.
The idea can be illustrated by an example.
Consider  the vertex which describes the fluctuations of the gauge vector
around the dipole string background 
expressed in the dipole string fields $X_{(0)}$, 
we can derive it from a generating functional for the dipole string as
\begin{align}
V_{(0)}(x;\epsilon,k)=
\epsilon_i \partial_x X_{(0)}^i(x,x) e^{ i k_j X_{(0)}^j(x,x)}
=
S_{(0) 0}(c,x) ~\epsilon_i 
\frac{ \stackrel{\leftarrow}{\partial} }{ \partial c_{1 i} } e^{ i k_j \frac{
    \stackrel{\leftarrow}{\partial} }{\partial c_{0 j} } }
\Big|_{c=0}
\end{align}
where we have introduced the generating vertex
\begin{align}
 S_{(0) 0}(c,x) 
&=
: e^{\sum_{k=0} ^\infty c_{k\, i}(x) ~\partial_x^k X^i_{(0)}(x,x)} : 
\end{align}
and the derivative $\partial_x$ acts on both the left moving and right
moving part. 
Then it is natural to assume that we can derive the emission vertex
for the same state  from the dicharged string as
\begin{align}
V(x;\epsilon,k)=
S_0(c,x) ~\epsilon_i 
\frac{ \stackrel{\leftarrow}{\partial} }{ \partial c_{1 i} } e^{ i k_j \frac{
    \stackrel{\leftarrow}{\partial} }{ \partial c_{0 j} } }
\Big|_{c=0}
\label{V-Tach-Using-c}
\end{align}
where $S_0(c,x)$ can be derived by regularizing the
analogous for the dicharged string of  the generating vertex 
$ S_{(0)  0}(c,x)$
as done in the previous section for the tachyonic vertex.
We write therefore
\begin{align}
[ S_0(c,x) ]_{p.s.}
&=
 e^{\sum_{k=0} ^\infty c^T_k ~\partial_x^k 
[ X^{(-)}(x e^{-\eta}, x e^{-\eta})+X^{(+)}(x, x) ]} 
\nonumber\\
&=
 e^{\sum_{k=0}^\infty c^T_k \frac{\uno+R_0}{2}~\partial^k
[ \Xln^{(-)}(x e^{-\eta})+\Xln^{(+)}(x)]
+ c_0^T x_0} 
\end{align}
where to write the last line we used the boundary conditions which
allow to write
\begin{align}
X(x,x)&= x_0+ \frac{\uno+R_0}{2} \Xln(x)
= x_0 + \cE_0^{-1} G  \Xln(x)
\end{align}
with $x\in \R^+$.
We can then deduce the final expression by first normal ordering and
then dividing by the regularization factor the previous expression
\begin{align}
S_0(c,x)
=&
\lim_{\eta\rightarrow 0^+} \cN_0(c,\eta) ~ [S_0(c,x)]_{p.s.}
\nonumber\\
=&
\exp\Big\{
-\alpha' \sum_{k,l=0}^\infty \sum_{c/\nu_c\ne0}
c_k^T\,\cE_0^{-1} G\,v_c~  v_c^\dagger\, G\cE_0^{-T}\, c_l ~
\partial^k_u|_{u=x}~\partial^l_v|_{v=x} ~\Delta_c(u/v)
\Big\}
\nonumber\\
&
\exp\Big\{
+\alpha' \sum_{l=0}^\infty \sum_{c/\nu_c\ne0}
c_0^T\,\cE_0^{-1} G\,v_c~  v_c^\dagger\, G\cE_0^{-T}\, c_l ~
\partial^l_x ~\log(x)
\Big\}
\nonumber\\
&
: e^{\sum_k c^T_k \cE_0^{-1} G ~\partial^k \Xln(x) + c_0^T x_0} :
\label{S-c-x}
\end{align}
where the last line can also be written as
$: e^{\sum_k c^T_k  ~\partial^k X(x,x)}: $ and
in the normal ordering the $x_0$ zero modes are not splitted.
We have also defined the function  
\begin{equation}
\Delta_a(u/v)
= \hat g_{-\nu_a}(u/v)- \hat g_0(u/v)
= \hat g_{-\nu_a}(u/v) - \ln(1-u/v)
\end{equation}
which near $u=v$ can be expanded as
\begin{equation}
\Delta_a(1+\eta)= 
\left\{ \begin{array}{c c}
\psi(\nu_a) -\psi(1) 
- (1 -\nu_a)\eta  +
\frac{(1-\nu_a)(2-\nu_a)}{4} \eta^2
+O(\eta^3) 
& \nu_a >0 
\\
\psi(1-|\nu_a|) -\psi(1) 
- |\nu_a|\eta  
+ \frac{|\nu_a|(1+|\nu_a|)}{4} \eta^2
+O(\eta^3) 
& \nu_a \le 0 
\end{array}
\right.
\end{equation}
\COMMENTO{
\begin{equation}
\Delta_a(1)= 
\left\{ \begin{array}{c c}
\psi(\nu_a) -\psi(1) & \nu_a >0 \\
\psi(1+\nu_a) -\psi(1) & \nu_a \le 0 
\end{array}
\right.
~~~~
\Delta_a'(1)= 
\left\{ \begin{array}{c c}
\nu_a-1 & \nu_a >0 \\
\nu_a & \nu_a \le 0 
\end{array}
\right.
\end{equation}
}

The true Sciuto - della Selva-Saito is then obtained 
by representing the algebra $[ c_{i k_1}, 
\frac{ \stackrel{\leftarrow}{\partial} }{ \partial  c_{j k_2}}]
= \delta_i^j~ \delta_{k_1, k_2}$
on the dipole string Fock space.
We introduce therefore the dipole string Fock space ${\cal H}_{(0)}$
then we make the identifications\footnote{
In the following we will also use $\alpha^i_n\equiv \sum_a v^i_a~
\alpha^a_n$ with $[\alpha^i_{n},\alpha^j_{m}]= \sum_a v^i_a (n+\nu_a)
v^j_{-a}\, \delta_{m+n,0} $. 
In the pure dipole string these $\alpha^i_{n (0)}$ 
satisfy  
$[\alpha^i_{(0) n},\alpha^j_{(0) m}]= n\, G^{i j}
 \delta_{m+n,0} $
and
are related to the usual
operators $\hat \alpha^i_n$, defined in the appendix \ref{app:Dipoleopen}, as 
$\alpha_{n (0)} = G^{-1} \cE_0 \hat \alpha_n$.
Moreover we find $[ x_0^i,\alpha_{n (0)}^j]= i \sqrt{2\alpha'}
(\cE_0^{-1})^{i j} \Rightarrow 
\alpha_{n (0)}^j \exp(i k_i x_0^i) |0\rangle=
\sqrt{2\alpha'} (\cE_0^{-1} k)^{j} \exp(i k_i x_0^i) |0\rangle
$ and similarly
$
\alpha_{n (\pi)}^j \exp(i k_i x_0^i) |0\rangle=
\sqrt{2\alpha'} (\cE_\pi^{-1} k)^{j} \exp(i k_i x_0^i) |0\rangle
$.
  }
\begin{align}
&
1 \rightarrow \langle x_{(0)}=0| \langle 0_{(0)} |
,~~~~
c_{k\, i} \rightarrow \frac{i}{\sqrt{2\alpha'}}
\frac{\alpha_{(0) k}^j}{k!} \cE_{0\, i j}
\nonumber\\
&
\frac{ \stackrel{\leftarrow}{\partial} }{ \partial  c_{i k}}
\rightarrow
-i \sqrt{2 \alpha'}\, (k-1)!\,(\cE_0 ^{-1} G)^i_j \alpha_{(0) -k}^j
\sim \partial^k X^{(-)i}_{(0)}(x,x) |_{x=0}
\end{align}
with $[\alpha^i_{(0) n},\alpha^j_{(0) m}]= n\,G^{i j}\, \delta_{m+n,0} $
so that after the substitution we get explicitly 
\begin{align}
\cS_0(x)
= 
\langle x_{(0)}=0| \langle 0_{(0)} |
&
\exp\Big\{\oh \sum_{k,l=0}^\infty \sum_{c/\nu_c\ne0}
\alpha_{(0) k}^{T}\,G \,v_c~  
v_c^\dagger\, G\, \alpha_{(0) l}
~\frac{\partial^k_u|_{u=x} }{k!}
~\frac{\partial^l_v|_{v=x} }{l!} ~\Delta_c(u/v)\Big\}
\nonumber\\
&
\exp\Big\{-\oh \sum_{l=0}^\infty \sum_{c/\nu_c\ne0}
\alpha_{(0) 0}^{T}\,G \,v_c~  
v_c^\dagger\, G\, \alpha_{(0) l} ~
\frac{\partial^l_x}{l!} ~\log(x)\Big\}
\nonumber\\
&
: \exp\Big\{i \frac{1}{\sqrt{2\alpha'}}\sum_{k=0}^\infty
  \frac{1}{k!}\alpha^{T}_{(0) k}
  ~\cE_0  ~\partial^k X(x,x)\Big\} 
:
\label{SDSzero}
\end{align}
In a similar way we can compute the  Sciuto della Selva-Saito
 vertex for the  emission of $\sigma=\pi$ dipole strings  
with the help of ($y<0$)
\begin{equation}
\nonumber\\
X(y,y)
= 
x_0 + \cE_\pi^{-1} G  \Xln(y)
- \sqrt{2\alpha' } \pi\, R_\pi \sum_{f/ \nu_f=0} v_f\, \alpha^f_0  
\end{equation}
as 
\begin{align}
\cS_\pi(y)
=& \langle x_{(\pi)}=0| \langle 0_{(\pi)} |
\exp\Big\{
\oh \sum_{k,l=0}^\infty \sum_{c/\nu_c\ne0}
\alpha_{(\pi)k}^{T}\,G\,v_c~  
v_c^\dagger\, G\, \alpha_{(\pi) l}
~\frac{\partial^k_u|_{u=y} }{k!}
~\frac{\partial^l_v|_{v=y} }{l!} ~\Delta_c(u/v)\Big\}
\nonumber\\
&
\exp\Big\{-\oh \sum_{l=0}^\infty \sum_{c/\nu_c\ne0}
\alpha_{(\pi)0 }^T\, G\,v_c~  
v_c^\dagger\, G\, \alpha_{(\pi) l} ~
\frac{\partial^l_y}{l!} ~\log(|y|)\Big\}
\nonumber\\
&
: e^{i \frac{1}{\sqrt{2\alpha'}}\sum_k
  \frac{1}{k!}\alpha_{(\pi) k }^T\,
  G\, \partial^k \Xln(y)} 
e^{i  \frac{1}{\sqrt{2\alpha'}} \alpha_{(\pi)0}^T\, \cE_\pi^T\,
(x_0 - \sqrt{2\alpha'} \pi R_0 v_e~\alpha^e_0 )}
:
\label{SDSpi}
\end{align}
where the auxiliary Fock space ${\cal H}_{(\pi)}$ is associated with the $\sigma=\pi$
dipole  Fock space and hence in principle different from the
$\sigma=0$ auxiliary Fock space.
The last line can also be rewritten as
$:e^{i \frac{1}{\sqrt{2\alpha'}}\sum_k  \frac{1}{k!}\alpha_{(\pi) k }^T
  ~\cE_\pi^{-1} ~\partial^k X(y,y)}: $ so that we immediately see that
$\Omega\,\cS_0(x)\,\Omega^{-1}= \tilde \cS_\pi(y)$ where $\tilde
\cS_\pi(y)$ is the SDS vertex associated to the emission from the
$\sigma=\pi$ boundary of the dicharged string with magnetic field
strength $(\tilde \cF_0=-\cF_\pi,\tilde \cF_\pi=-\cF_0)$.

The corresponding closed string emission vertex can be written as
generating function  as
\begin{align}
\cS(c_L,c_R,z,\bar z)
=&
\exp\Big\{
-\alpha' \sum_{k,l=0}^\infty \sum_{c/\nu_c\ne0}
c_{L\,k}^T\,v_c~  v_c^\dagger\, c_{L\,l} ~
\partial^k_u|_{u=z}~\partial^l_v|_{v=z} ~\Delta_c(u/v)
\Big\}
\nonumber\\
&
\exp\Big\{
+\alpha' \sum_{l=0}^\infty \sum_{c/\nu_c\ne0}
c_{L\,0}^T\,v_c~  v_c^\dagger\, c_{L\,l} ~
\partial^l_z ~\log(z)
\Big\}
\nonumber\\
&
\exp\Big\{
-\alpha' \sum_{k,l=0}^\infty \sum_{c/\nu_c\ne0}
c_{R\,k}^T\,R_0\,v_c~  v_c^\dagger\,R_0^T\, c_{R\,l} ~
\bar\partial^k_u|_{u=\bar z}~\bar\partial^l_v|_{v=\bar z} ~\Delta_c(u/v)
\Big\}
\nonumber\\
&
\exp\Big\{
+\alpha' \sum_{l=0}^\infty \sum_{c/\nu_c\ne0}
c_{R\,0}^T\,R_0\,v_c~  v_c^\dagger\, R_0\,c_{R\,l} ~
\bar\partial^l_{\bar z} ~\log(\bar z)
\Big\}
\nonumber\\
&
: e^{\sum_k c^T_{L\, k} ~\partial^k \Xln(z) +
  c_{L\,0}^T\,G^{-1}\cE_\pi\, x_0} :
~~
: e^{\sum_k c^T_{R\, k} ~\bar \partial^k \Xrn(\bar z) +
  c_{R\,0}^T\,G^{-1}\cE_\pi^T\, x_0} :
\label{SDSclosed-c}
\end{align}
up to a cocycle discussed and determined for the closed string tachyon
in appedix \ref{AppDetailsVertexes}.
Upon the identifications with the usual closed string operators 
\begin{align}
&
1 \rightarrow \langle x_{L}=0| \langle 0_{L} |
,~~~~
c_{L\,k\, i} \rightarrow \frac{i}{\sqrt{2\alpha'}}
\frac{\alpha_{L\, k}^j}{k!} G_{ i j}
\nonumber\\
&
\frac{ \stackrel{\leftarrow}{\partial} }{ \partial  c_{L\,i\, k}}
\rightarrow
-i \sqrt{2 \alpha'}\, (k-1)!\, \alpha_{L\, -k}^i
\sim \partial^k X^{(-)i}_{L}(z) |_{z=0}
\end{align}
and analogously for the right modes the true SDS reads
\begin{align}
\cS(z,\bar z)
=
\langle x_{L}=0| \langle 0_{L} |
&
\exp\Big\{\oh \sum_{k,l=0}^\infty \sum_{c/\nu_c\ne0}
\alpha_{L\, k}^{T}\,G \,v_c~  
v_c^\dagger\, G\, \alpha_{L\, l}
~\frac{\partial^k_u|_{u=z} }{k!}
~\frac{\partial^l_v|_{v=z} }{l!} ~\Delta_c(u/v)\Big\}
\nonumber\\
&
\exp\Big\{-\oh \sum_{l=0}^\infty \sum_{c/\nu_c\ne0}
\alpha_{L\, 0}^{T}\,G \,v_c~  
v_c^\dagger\, G\, \alpha_{L\, l} ~
\frac{\partial^l_z}{l!} ~\log(z)\Big\}
\nonumber\\
&
: \exp\Big\{
i \frac{1}{\sqrt{2\alpha'}}\sum_{k=0}^\infty
  \frac{1}{k!}\alpha^{T}_{L\, k}
   ~\partial^k \Xln(z)
+
i \frac{1}{\sqrt{2\alpha'}} \alpha^{T}_{L\, 0} \cE_\pi x_0
\Big\} 
:
\nonumber\\
\langle x_{R}=0| \langle 0_{R} |
&
\exp\Big\{\oh \sum_{k,l=0}^\infty \sum_{c/\nu_c\ne0}
\alpha_{R\, k}^{T}\,G R_0 \,v_c~  
v_c^\dagger\, R_0^T G\, \alpha_{R\, l}
~\frac{\bar\partial^k_u|_{u= \bar z} }{k!}
~\frac{\bar\partial^l_v|_{v=\bar z} }{l!} ~\Delta_c(u/v)\Big\}
\nonumber\\
&
\exp\Big\{-\oh \sum_{l=0}^\infty \sum_{c/\nu_c\ne0}
\alpha_{R\, 0}^{T}\,G R_0\,v_c~  
v_c^\dagger\, R_0^T G\, \alpha_{R\, l} ~
\frac{\bar \partial^l_{\bar z}}{l!} ~\log(\bar z)\Big\}
\nonumber\\
&
: \exp\Big\{
i \frac{1}{\sqrt{2\alpha'}}\sum_{k=0}^\infty
  \frac{1}{k!}\alpha^{T}_{R\, k}
   ~\bar\partial^k \Xrn(\bar z)
+
i \frac{1}{\sqrt{2\alpha'}} \alpha^{T}_{R\, 0} \cE_\pi^T x_0
\Big\} 
:
\label{SDSclosed}
\end{align}
It is worth stressing that the splitting of zero modes $x_0$ in left
$x_{0 L}= G^{-1} \cE_\pi x_0 $
and right $x_{0 R}= G^{-1} \cE_\pi^T x_0 $
parts
is dictated by the request that the OPE of two of the
previous vertexes reproduces the usual OPE (see appendix
\ref{AppDetailsVertexes} ).

\subsection{Examples of vertexes}
\label{ExamplesV}
We can now put at work the results of the previous section.
In this section we write only the part of the vertex in the twisted directions. 
We start with simplest case, i.e. the tachyonic vertex from the
$\sigma=0$ boundary
\begin{align}
V_{(0) T_0}(x,k) = : e^{i k_i X^i_{(0)}(x,x)} :
~\Rightarrow~
V_{(0) T_0}(0,k) |0_{(0)}\rangle = |k_{(0)}\rangle
\end{align}
so we can compute the emission vertex of the same tachyon from the dicharged
string as
\begin{align}
V_{ T_0}(x,k) =& \cS_{0}(x)|k_{(0)}\rangle
\nonumber\\
=&
e^{\alpha' k_i\, (\cE_0^{-1} G\,v_c)^i~  (v_c^\dagger\, G\cE_0^{-T})^j\, k_j~
  \Delta_c(1)}
\nonumber\\
&
x^{ -\alpha' k_i 
\,(\cE_0^{-1} G\,v_c)^i~  (v_c^\dagger\, G\cE_0^{-T})^j\,
k_j
}
\nonumber\\
&
: e^{i k_i X^i(x,x)} :
\end{align}
which reduces to eq. (\ref{VT0}) when we consider the dicharged string
on $\R^2$.
In a similar way we get the tachyonic vertex from the
$\sigma=\pi$ boundary
\begin{align}
V_{ T_\pi}(y,k) =& \cS_{\pi}(y)|k_{(\pi)}\rangle
\nonumber\\
=&
e^{\alpha' k_i\, (\cE_\pi^{-1} G\,v_c)^i~  (v_c^\dagger\, G\cE_\pi^{-T})^j\, k_j~
  \Delta_c(1)}
\nonumber\\
&
|y|^{ -\alpha' k_i 
\,(\cE_\pi^{-1} G\,v_c)^i~  (v_c^\dagger\, G\cE_\pi^{-T})^j\,
k_j
}
\nonumber\\
&
: e^{i k_i X^i(y,y)} :
\end{align}
which differs from the  $\sigma=0$ one also because it depends  on
$\cE_\pi$ in stead of $\cE_0$.

As a second example we consider the fluctuations of the gauge fields.
This result will be used later when we compute the instanton form
factor. The dipole state associated with the emission of this
fluctuation is 
\begin{align}
V_{(0) A}(x,\epsilon,k) = : \epsilon_j \partial X^j(x,x)\,e^{i k_i X^i_{(0)}(x,x)} :
~\Rightarrow~
V_{(0) A}(0,\epsilon,k) |0_{(0)}\rangle = 
- i\sqrt{2\alpha' } \epsilon_j (\cE_0^{-1} G)^j_{~ l} \alpha^l_{(0)-1}|k_{(0)}\rangle
\end{align}
so we get the emission vertex  from the dicharged string as
\begin{align}
V_{A}(x,\epsilon,k) =& \cS_{0}(x)
\Big[- i\sqrt{2\alpha' } \epsilon_j 
(\cE_0^{-1} G)^j_{~ l}  \alpha^l_{(0)-1}|k_{(0)}\rangle\Big]
\nonumber\\
=&
e^{\alpha' k_i\, [\cE_0^{-1} G\,v_c~  v_c^\dagger\, G\cE_0^{-T}]^{i j}\, k_j
  ~\Delta_c(1)}
\nonumber\\
&
x^{ -\alpha' k_i 
[
\,\cE_0^{-1} G\,v_c~  v_c^\dagger\, G\cE_0^{-T}\,
]^{i j}\, 
k_j
}
\nonumber\\
&
:\Bigg[
\epsilon_j \partial X^j(x,x)
+ 
i \alpha' 
\frac{2( \theta(-\nu_c) + \nu_c)}{x}
k_i\,  [\cE_0^{-1} G\, v_c ~v_c^\dagger\, G\cE_0^{-T}]^{i j}\, \epsilon_j
\Bigg] 
e^{i k^T X(x,x)} :
\end{align}
 where $\theta(x)=1$ for $x>0$ and $0$ otherwise is the Heaviside function.
Obviously the same result can be obtained computing
(\ref{V-Tach-Using-c}) using the $\cS(c,x)$ expression given in
eq. (\ref{S-c-x}). 
The previous vertex can also be rewritten in a different way by
splitting the last term into symmetric and antisymmetric part as
\begin{align}
V_{A}(x,\epsilon,k) =& 
e^{\alpha' k^T\, \cE_0^{-1} G\,v_c~  v_c^\dagger\, G\cE_0^{-T}\, k~
  \Delta_c(1)}
\nonumber\\
&
x^{ -\alpha' k^T 
\,\cE_0^{-1} G\,v_c~  v_c^\dagger\, G\cE_0^{-T}\,
 k
}
\nonumber\\
&
:\Bigg[
\epsilon^T \partial X(x,x)
+
i \alpha' 
\frac{1}{x}
k^T\,  \cE_0^{-1} G\, v_c ~
v_c^\dagger\, G\cE_0^{-T}\, \epsilon
+
 \alpha' 
\frac{1}{x}
k^T\, \Theta_{vect}\,  \epsilon
\Bigg] 
e^{i k^T X(x,x)} :
\end{align}
where $\Theta_{vect}= -\Theta^T_{vect}= 
i\,\sum_{c>0}(1 - 2 |\nu_c|) 
\cE_0^{-1} G\, 
(v_c ~ v_c^\dagger - v_{-c} v_{-c}^\dagger )\, 
G\cE_0^{-T}$ 
and we cannot use the spectral decomposition of $G^{-1}$ in the
symmetric part since
$\{v_c\}$ do not generically form a basis.

Another interesting example is the energy momentum tensor. It is a
descendant of the unity $1_{(0)}$ of the dipole string and therefore
it can be computed as
\begin{align}
T_{(0) 1_0}(x) 
&= 
-\frac{1}{4\alpha' } :\partial_x X_{L (0)}^T\, G\, \partial_x X_{L (0)}:
\nonumber\\
&\Rightarrow
T_{(0) 1_0}(0) |0\rangle_{(0)} =  \oh \alpha_{(0) -1} ^T\, G\, \alpha_{(0)  -1}
|0\rangle_{(0)}
\nonumber\\
&\Rightarrow
T_{1_0}(x) =  \cS_0(x)~
\oh  \alpha_{(0) -1} ^T\, G\, \alpha_{(0) -1}|0\rangle_{(0)}
\nonumber\\
&~~~~=
-\frac{1}{4\alpha' } \partial_x X_{L }^T\, G\, \partial_x X_{L}
+\oh \sum_{c/ \nu_c \ne 0} \partial_u|_{u=x} \partial_v|_{v=x} \Delta_c(u/v)
\end{align}
On the other side the  energy momentum tensor of the dicharged string 
$T(z)$ has exactly the same OPEs with all the dipole string vertexes
in dicharged string formalism and therefore we can identify on the boundary
$T(x)=T_{(0) 1_0}(x)$ and analogously for the $\sigma=\pi$ dipole string 
$T(y)=T_{(\pi) 1_\pi}(y)$ ($y<0$).
Therefore we recover the dicharged string energy momentum
(\ref{dichargedT}) when we rewrite the second addend as
$
-\frac{1}{2 x^2}\sum_{c>0}(
\Delta'_c(1)+ \Delta''_c(1)
+\Delta'_{-c}(1)+ \Delta''_{-c}(1)
)
= 
+\frac{1}{2 x^2}\sum_{c>0} \nu_c(1-\nu_c)
$.

\subsection{Consistency conditions for the vertexes.}
\label{sect:consistency}
We have constructed the SDS vertexes and some explicit examples of
vertexes which can be computed from it. 
We have now ready to state the consistency conditions we expect to be
satisfied by the proposed vertexes in order to be identified with the
proper emission vertexes of the dipole string states from the dicharged
string:
\begin{enumerate}
\item
\label{conf-prop-in}
open string vertexes have well defined conformal transformations;
\item
\label{conf-prop-fin}
closed string vertexes has well defined conformal transformations;
\item
\label{comm-prop-in}
 open string vertexes on the opposite boundaries commute;
\item
\label{comm-prop-fin}
open string emission vertexes  commutes with the
closed string vertexes;
\item
\label{ope-prop-in}
the OPE of two open string vertexes for the emission of dipole
states from the $\sigma=0$ boundary of the dicharged string
must have the same coefficients of the OPE of the corresponding open
string vertexes for the emission of the same dipole
states from the $\sigma=0$ boundary of the dipole string when we
properly map the vertexes;
\item
the same is true for the emission of dipole states from the
$\sigma=\pi$ boundary;
\item
\label{ope-prop-fin}
in a similar way the OPE of two closed string vertexes in open string formalism
must reproduce the result in the closed string formalism.
\end{enumerate}
Details on how to check the previous constraints are given in appendix
\ref{AppDetailsVertexes}.
Here we limit ourselves to some comments.
\begin{itemize}
\item 
Constraints (\ref{conf-prop-in}-\ref{conf-prop-fin}) are actually a
consequence of  (\ref{ope-prop-in}-\ref{ope-prop-fin}) since we have
shown that the dicharged energy-momentum tensor can be identified with
the dipole energy-momentum tensor.
In any case because in the generic OPE 
\begin{equation}
T(z) V(w) =
 \frac{\Delta ~V(w)}{(z-w)^2}+
\frac{\partial V(w)}{(z-w)}+ reg
\end{equation}
there is the derivative of $V$ 
from  constraints (\ref{conf-prop-in}-\ref{conf-prop-fin})
 we can test  the presence of terms
like $x^{-\Delta}$ in the vertex nevertheless the purely numerical
normalization factors like $e^{\delta(\epsilon) \Delta_0(k)}$ cannot be tested. 
\item
The constraints (\ref{comm-prop-in}-\ref{comm-prop-fin}) 
check the operatorial structure of vertexes only. 
\item
The constraints (\ref{ope-prop-in}-\ref{ope-prop-fin})  
check both the operatorial and the non-operatorial structure
of the vertexes.
\end{itemize}


\section{Stringy form factors}
\label{sect:formfactors}
We are now ready to use the previously developed machinery to do some
amplitude computations.
\subsection{Tachyonic form factor on $\R^2$}
As a warming up we consider the tachyonic profile of the
$D25/D25'$ bound state where the first $D25$ has a background field strength
$(F_0)_{1 2}$ switched on and the second $(F_\pi)_{1 2}$.
The amplitude we consider can be written in the usual CFT formalism as
\begin{align}
\cA
&=
\cC_0(\nu) \cN_0(\nu)^2 \cN_0(F_0)
~\langle 
V_{T25'/25}(x_1; q_\mu,\lambda)~
V_{T25}(x_2; k_M)
V_{T25/25'}(x_3; p_\mu,\kappa)
~\rangle
\end{align}
where $\cC_0(\nu)$ is the normalization of the mixed amplitude,
$\cN_0(\nu)$ is the normalization of the dicharged vertex,
$\cN_0(F_0)$ is the normalization of the dipole vertex and
$\mu=0,3,\dots D-1$, $M=(\mu,i)=0,\dots D-1$. 
The dicharged vertex for the tachyon with momentum $p_\mu$
($\mu\ne1,2$), polaritazion $t(p_\mu,\kappa)$  
is given by
\begin{align}
V_{T25/25'}(x; p_\mu,\kappa)
=
t(p_\mu,\kappa)~
c(x)\, \Delta_{\nu,\kappa}(x)\, e^{i p_\mu X^\mu(x,x)}
\end{align}
with $\alpha' p_\mu^2+ \nu(1-\nu)=1$ and
where we have introduced the family of twist fields
$\Delta_{\nu,\kappa}(x)$ parametrized by $\kappa$. The reason of
introducing this family is that the vacuum is degenerate and labeled
by $\kappa$ (which we can identify with the ``momentum'' in direction
2) as in eq. (\ref{spectrum}) but it is not possible to
account for this degeneracy introducing a factor $e^{i \kappa X^2}$ in
the vertex because it would change the conformal dimension.
The dipole vertex with polarization $T(k_M)$ and momentum $k_M=(k_\mu,k_i)$
($M=0,\dots D-1$, $i=1,2$) is given as usual by
\begin{align}
V_{T25}(x; k_M)
=
T(k_M)~
c(x)\, e^{i k_M X^M(x,x)}
\end{align}
with  $\alpha'( k_\mu^2+  \cG_0^{i j} k_i k_j) =1$.
The computation is straightforward for all correlators but 
the three point correlator
\begin{align}
\langle
\Delta_{-\nu,\lambda}(x_1)
~ e^{i k_i  X^i(x_2,x_2)}
~\Delta_{\nu,\kappa}(x_3)
\rangle
&=
\frac{\cC}{x_{12}^{\Delta(k_i)} x_{13}^{\nu(1-\nu)-\Delta(k_i)} 
x_{23}^{\nu(1-\nu)-\Delta(k_i)}}
\end{align}
which is completely fixed by conformal invariance up to the constant
$\cC$. In the previous expression $\Delta(k_i)=\alpha' \cG_0^{i j} k_i
k_j$ is the  conformal dimension of $e^{i k_i X^i(x_2,x_2)}$.
To fix the constant $\cC$ we take the limit $x_3\rightarrow 0$,
$x_1\rightarrow \infty$ and we get
\begin{align}
\lim_{x_3\rightarrow 0, x_1\rightarrow \infty}
&
x_1^{\nu(1-\nu)}
\langle
\Delta_{-\nu,\lambda}(x_1)
~ e^{i k_i  X^i(x_2,x_2)}
~\Delta_{\nu,\kappa}(x_3)
\rangle
= \frac{\cC}{x_2^{\Delta(k_i)}}
\nonumber\\
&=
\langle 0_\nu, -\lambda|
e^{-\ce \Delta(k_i)} x_2^{-\Delta(k_i)} e^{i k_i X^i_{(\nu)}(x_2,x_2)}
| 0_\nu,\kappa\rangle
\nonumber\\
&=
e^{-\ce \Delta(k_i)}\, x_2^{-\Delta(k_i)}\, 
\langle  -\lambda| e^{i k_i x^i_0} |\kappa\rangle
\nonumber\\
&=
e^{-\ce \Delta(k_i)}\, x_2^{-\Delta(k_i)}\, 
e^{-i \pi \alpha'\, k_1 k_2\, \Theta_{dicharged}^{12}}\,
e^{i k_1 \kappa}
2\pi \delta(k_2+\kappa+\lambda)
\end{align}
with $ \ce= -(\Delta_{c=1}(1)+\Delta_{c=-1}(1))$ and
where the result of the last line is due to the fact that the
correlator boils down to a quantum mechanical computation for the
Landau level theory with
$\Theta_{dicharged}^{12}=((\cF_\pi-\cF_0)^{-1})^{12}$.
This result can seem quite strange since it involves $k_1$ and $k_2$ in an
asymmetric way. The reason is that we started with the vacuum annihilated
by $x^1_0$ but we could as well have started with a vacuum annihilated
by any linear combination of $x^1_0$ and $x^2_0$. In particular
choosing the vacuum annihilated by  $x^2_0$ would reverse the role of
the two directions.

Putting all together we get the amplitude
\begin{align}
\cA
=&
\cC_0(\nu) \cN_0(\nu)^2 
\cN_0(F_0)
\,(2\pi)^{D-2} \delta^{D-2}(p_\mu+k_\mu+q_\mu)
\,2\pi \delta(k_2+\kappa+\lambda)
\nonumber\\
&
t(p_\mu,\kappa)~t(q_\mu,\lambda)~T(k_M)~
\,e^{-i \pi \alpha'\, k_1 k_2\, \Theta_{dicharged}^{12}}\,
e^{i k_1 \kappa}
\,e^{-\ce \alpha' \cG_0^{i j} k_i k_j}
\end{align}
where the last term can be interpreted as form factor of the twisted
matter. One could doubt about this interpretation because if we had
computed the amplitude with two dipole tachyons with momenta $k$ and
$l$ each vertex would yield $e^{-\ce \alpha' \cG_0^{i j} k_i k_j}$
and $e^{-\ce \alpha' \cG_0^{i j} kl_i l_j}$ respectively which is
not the expected factor $e^{-\ce \alpha' \cG_0^{i j} (k+l)_i (k+l)_j}$.
Nevertheless this factor is what one gets when factorizing in the
dipole string channel because of OPEs.

Let us now discuss the tachyonic profile of a $D25/D23$ system.
We start considering 
the $D25$ with a vanishing background field strength $(F_0)_{12}$
and with zero Kalb-Ramond $B_{12}=0$ so  that $\nu=\oh$.
Nevertheless this system cannot be obtained  
taking the $(F_\pi)_{1 2} \rightarrow \infty $  limit rigorously.
In fact taking naively this limit 
we expect the same infinite degeneration of the
$D25/D25'$ case and this expectation is generically true 
unless we both choose $\nu=\oh$ and consider 
the $D25/D23$ configuration from the beginning in a compact space on
which we take the decompactification limit.
To understand this point it is necessary to start with a compactified version
of the theory on $\R^{D-2}\times T^2$ and 
either construct explicitly the vacuum or more
intuitively take a T-duality and end with a $D24/D24'$ system. 
In this T-dual version we expect that the two $D24$ meet only once, i.e. that
the first Chern class be $c_1=1$ but 
$c_1=1$ of more generically $c_1$ finite 
cannot be true on a space with arbitrary large volume.
A naive way of seeing this is to notice that 
$c_1 \propto (F_0 -F_\pi)_{12} ~vol(T^2)$ so that $c_1$ cannot stay
finite when $ vol(T^2) \rightarrow \infty$ with non vanishing $F_0 -F_\pi$.
Another way of getting the same result in a slightly more general case
with non vanishing $B_{12}$
is to realize that 
if we require $\nu=\oh$ we must have $(F_0-B)_{1 2}(F_\pi-B)_{12}=-\det G$
which allows a solution for $B_{12}$ only when $(F_0-F_\pi)_{1 2}^2 > 4\, \det G$.
This last equation can be rewritten as $ \frac{c_1^2}{(N_0 N_\pi)^2}  > 4\, \det
G$ ($N_0,N_\pi$ arbitrary number of $D24$ and $D24'$ which can be
obtained by T-duality) in a compact space so we
cannot take the decompactification limit unless we take
$c_1\rightarrow \infty$ at the same time. 
This is exactly the same result we get by taking the  $(F_\pi)_{1 2}
\rightarrow \infty $ naively.
The only way to get the vacuum degeneration $c_1=1$ (or finite) in the 
non compact case is to start already from  a $D25/D23$ system.
Even if we start from a system $D25/D23$ with
$B_{12}=0$ 
and a
non vanishing background field strength $(F_0)_{12}$ on a compact
space and we perform a T-duality to get a $D25/D25'$ system 
the volume
of the $T^2$ on which this latter system lives is bounded as $\det G<
\frac{1}{4}$ and so we cannot take the decompactification limit.
Explained this point if we assume $c_1=1$ and to take the
decompactification limit in the proper way then the zero modes
correlator gives
\begin{equation}
\langle\lambda=0,y_0| e^{i k_i x^i_0} | \kappa=0,y_0\rangle
=e^{i k_i y^i_0}
\end{equation}
where $\lambda=\kappa=0$ is the only possible value of the
vacuum degeneration label and $y_0$ is the Wilson line which appears
as background only for compact spaces.
Notice that the previous state with $c_1=1$ (or finite) 
is the analogous of a momentum eigenstate
with discrete momentum while the one entering the spectrum in
eq. (\ref{spectrum}) is analogous  to one with continuum momentum and
so the two are connected as
\begin{equation}
|\lambda\rangle_{continuum}=  \sqrt{2 \pi c_1} |\lambda\rangle_{discrete}
\end{equation}
where $c_1$ plays the role of the radius.
The same kind of issue in defining the lower dimensional brane vacuum
is present also for the dipole string
(\cite{Pesando:2003ww},\cite{Pesando:1999hm}). 
Finally we can write the amplitude for the $D25/D23$ case as
\begin{align}
\cA
&=
\cC_0(\nu) \cN_0(\nu)^2 \cN_0(F_0)
\,(2\pi)^{D-2} \delta^{D-2}(p_\mu+k_\mu+q_\mu)
t(p_\mu)~t(q_\mu)~T(k_M)\,
e^{i k_i y^i_0}
\,e^{-\ce \alpha' \cG_0^{i j} k_i k_j}
\end{align}
with $\pi \nu= \oh \pi - arctg\, (\ff_0/\sqrt{\det G_{T^2}}) $ and where the dicharged
tachyon polarizations now depend only on the momenta and not on the
parameter labeling the infinite degeneracy of the Landau levels in the
non compact case.

\subsection{Electromagnetic form factor on $\R^2$}

Let us now consider the electromagnetic profile of the $D25/D25'$
matter which can be computed by the correlator
\begin{align}
\cA
&=
\cC_0(\nu) \cN_0(\nu)^2 \cN_0(F_0)
~\langle V_{T25/25'}(x_1; q_\mu,\lambda)~
V_{A25}(x_2; k_M,\epsilon_M)
V_{T25'/25}(x_3; p_\mu,\kappa)~
\rangle
\end{align}
where the dipole vertex is given as usual by
\begin{align}
V_{A25}(x; k_M, \epsilon)
=
\epsilon_N(k_M)~
c(x)\, \partial X^N(x,x)\,e^{i k_M X^M(x,x)}
\end{align}
with  $\alpha'( k_\mu^2+  \cG_0^{i j} k_i k_j) = \epsilon_\mu k^\mu +
\epsilon_i \cG_0^{i j} k_j =0$.
The computation is again straightforward for all correlators but 
the three points correlator
\begin{align}
\langle
\Delta_{-\nu,\lambda}(x_1)~ 
\partial X^j(x_2,x_2)e^{i k_i  X^i(x_2,x_2)}
~\Delta_{\nu,\kappa}(x_3)
\rangle
&=
\frac{\cC(x_1,x_2,x_3)
}{
x_{12}^{\Delta(k_i)+1} x_{13}^{\nu(1-\nu)-\Delta(k_i)-1} 
x_{23}^{\Delta(k_i)+1}
}
\nonumber\\
=
\frac{
-i \alpha' \left(
\frac{x_{23}-x_{12}}{x_{13}}\, \cG_0^{j l} 
-i
\Theta_{vect}^{j l}
\right)\,k_l
}{
x_{12}^{\Delta(k_i)+1} 
x_{13}^{\nu(1-\nu)-\Delta(k_i)-1} 
x_{23}^{\Delta(k_i)+1}
}
&
e^{-\ce \Delta(k_i)}\, 
e^{-i \pi \alpha'\, k_1 k_2\, \Theta_{dicharged}^{12}}\,
e^{i k_1 \kappa}\,
2\pi \delta(k_2+\kappa+\lambda)
\end{align}
with
$\Theta_{vect}
=
(1-2 |\nu_1|)
\cE_0^{-1} G  \left( \begin{array}{c c} 0 & 1 \\ -1 & 0 \end{array} \right)
G \cE_0^{-T}
$.
The previous correlator is  not fixed in the functional dependence on
$x$ by conformal invariance since
$\partial X^j(x_2,x_2)e^{i k_i  X^i(x_2,x_2)}$ is not a good conformal
operator because of the cubic pole with the energy-momentum tensor.
We can fix this correlator by first taking the limit
\begin{align}
\lim_{x_3\rightarrow 0, x_1\rightarrow \infty}
&
x_1^{\nu(1-\nu)}
\langle
\Delta_{-\nu,\lambda}(x_1)~ 
\partial X^j(x_2,x_2) e^{i k_i  X^i(x_2,x_2)}
~\Delta_{\nu,\kappa}(x_3)\rangle
= 
\nonumber\\
&=
\langle 0_\nu, -\lambda|
e^{-\ce \Delta(k_i)} x_2^{-\Delta(k_i)} e^{i k_i  X^i_{(\nu)}(x_2,x_2)}
\Big(
\partial X^j_{(\nu)}(x_2,x_2)
+ i\alpha' \frac{\cG_0^{j l} + i \Theta^{j l}_{vect} }{x_2}  k_l
\Big)
| 0_\nu,\kappa\rangle
\nonumber\\
&=
e^{-\ce \Delta(k_i)}\, x_2^{-\Delta(k_i)-1}\, 
\alpha' \left( i \cG_0^{j l} -\Theta^{j l}_{vect} \right) k_l\,
\langle  -\lambda| e^{i k_i x^i_0} |\kappa\rangle
\nonumber\\
&=
e^{-\ce \Delta(k_i)}\, x_2^{-\Delta(k_i)-1}\, 
e^{-i \pi \alpha'\, k_1 k_2\, \Theta_{dicharged}^{12}}\,
\alpha' \left( i \cG_0^{j l} -\Theta^{j l}_{vect} \right) k_l\,
e^{i k_1 \kappa}\,
2\pi \delta(k_2+\kappa+\lambda)
\end{align}
then computing the amplitude in this limit
\begin{align}
&\lim_{x_1\rightarrow \infty, x_3\rightarrow 0} x_2^2 \cA 
=
\cC_0(\nu) \cN_0(\nu)^2 \cN_0(F_0)
\,(2\pi)^{D-2} \delta^{D-2}(p_\mu+k_\mu+q_\mu)
\,2\pi \delta(k_2+\kappa+\lambda)
\nonumber\\
&
(-i \alpha')
t(p_\mu,\kappa)~t(q_\mu,\lambda)\,
\left[\epsilon_\mu(k_M) (p^\mu-q^\mu)-i \epsilon_j(k_M) \Theta^{j l}_{vect}  k_l\right]
\,e^{-i \pi \alpha'\, k_1 k_2\, \Theta_{dicharged}^{12}}\,
e^{i k_1 \kappa}
\,e^{-\ce \alpha' \cG_0^{i j} k_i k_j}
\end{align}
and finally reconstructing the correlator such that it reproduces this
amplitude without taking the ${x_3\rightarrow 0, x_1\rightarrow
  \infty}$ limit.

Again the $D25/D23$ amplitude is slightly different because of zero
modes and reads
\begin{align}
\cA
&=
\cC_0(\nu) \cN_0(\nu)^2 \cN_0(F_0)
\,(2\pi)^{D-2} \delta^{D-2}(p_\mu+k_\mu+q_\mu)
\nonumber\\
&
(-i \alpha')
t(p_\mu)~t(q_\mu)\,
\left[\epsilon_\mu(k_M)\, (p^\mu-q^\mu)-i \epsilon_j(k_M)\, \Theta^{j l}_{vect}  k_l\right]
e^{i k_i y^i_0}
\,e^{-\ce \alpha' \cG_0^{i j} k_i k_j}
\end{align}

\subsection{Electromagnetic form factor on $\R^d$}
We want to to generalize the previous computation to $\R^d$ with
$d_K= \mbox{dim}~ ker(\Delta\cF)$. The first issue is to write the
tachyonic dicharged vertex. It is not completely trivial since  we can
immediately write
\begin{align}
V_{T25/25'}(x; p_\mu,p_f,\kappa)
=
t(p_\mu,p_f,\kappa)~
c(x)\, \Delta_{\nu,\kappa}(x)\, e^{i p_\mu X^\mu(x,x)+i p_f X^f(x,x)}
\label{VT-general}
\end{align}
but then we have to specify what $X^f(x,x)$ is. There are at least three
possibilities $v_f^\dagger G X(x,x)$, $v_f^\dagger \cE_0 X(x,x)$ and
$v_f^\dagger \cE_\pi X(x,x)$.  By comparing with the state
$|p_f\rangle=e^{i p_f x^f_0} |0\rangle$ and its dimension 
$L_0 |p_f\rangle= 2\alpha'\, p_f p_{-f} |p_f\rangle$, it turns out
that
\begin{equation}
X^f(x,x)= v_f^\dagger \cE_\pi X(x,x).
\end{equation}
In the previous vertex (\ref{VT-general}) $\nu$ and $\kappa$ are now  
$
\oh \mbox{dim}\, ker( \Delta \cF)^\perp= \oh(d-d_K)
$
dimensional vectors which label the Landau levels ``frequencies''  and
their degeneracies.
It is then easy to compute
\begin{align}
\cA
&=
\cC_0(\nu) \cN_0(\nu)^2 \cN_0(F_0)
\,(2\pi)^{D-d} \delta^{D-d}(p_\mu+k_\mu+q_\mu)
\,(2\pi)^{d_K} \delta^{d_K}(p_f+k_f+q_f)
\,\langle-\lambda| e^{i k_c x^c_0} |\mu \rangle
\nonumber\\
&
(-i \alpha')
t(p)~t(q)\,
\left[\epsilon_\mu(k) (p^\mu-q^\mu)+ \epsilon_f(k) (p_{-f}-q_{-f})
-i \epsilon_j(k) \Theta^{j l}_{vect}  k_l\right]
\,e^{- \alpha' \sum_{c>0} R^2(\nu_c) \hat k_c \hat k_{-c}}
\label{D25DpGeneral}
\end{align}
where we have defined
$
k_f= k^T \cE_\pi^{-1}G v_f= \hat k_f=  k^T \cE_0^{-1} G v_f
$
and
$
k_c= k^T \cE_\pi^{-1}G v_c\ne \hat k_c=  k^T \cE_0^{-1} G v_c
$
which are two possible ways of projecting the momentum $k_i$
($i=1,\dots d$) along the directions with $\nu_f=0$  and $\nu_c\ne0$.
We also explicitly have
$
-i \epsilon_j \Theta^{j l}_{vect}  k_l= 
\sum_{c>0} (1-2 \nu_c) (\hat\epsilon_c\, \hat k_{-c} -\hat k_c\, \hat \epsilon_{-c}) 
$.
Moreover we have left not evaluated the quantum mechanical correlator
$
\langle-\lambda| e^{i k_c x^c_0} |\mu \rangle
$
because its value depends whether we consider a $D25/D25'$ system or  a
$D25/D p$ one since for any dimension less than 25 we gain a Wilson
line and loose an infinite degeneracy if there is already a zero eigenvalue.
Also $t(p)$ and $t(q)$  have a different dependence on parameters
whether we consider  $D25/D25'$ or $D25/D p$ as discussed above.

The previous amplitude can be computed without taking the limit using
the correlator
\begin{align}
&
\langle
\Delta_{-\nu,\lambda}(x_1)~ 
\partial X^j(x_2,x_2)e^{i k_i  X^i(x_2,x_2)}
~\Delta_{\nu,\kappa}(x_3)
\rangle
=
\nonumber\\
&~~~
=
\frac{
-i \alpha' \left(
2 \frac{x_{12}}{x_{13}} (\cE_0^{-1} C v_f)^i\, p_{-f}
-2 \frac{x_{23}}{x_{13}} (\cE_0^{-1} C v_f)^i\, q_{-f}
+\frac{x_{23}-x_{12}}{x_{13}}\,  (\cE_0^{-1} C v_c)^i\, \hat k_{-c}
-i \Theta_{vect}^{j l} k_l\right)
}{
x_{12}^{\alpha'(q_f q_{-f} -p_f p_{-f}) +\Delta(k_i)+1} 
x_{13}^{\sum_{c>0}\nu_c(1-\nu_c)+\alpha'(q_f q_{-f} +p_f p_{-f})-\Delta(k_i)-1} 
x_{23}^{\alpha'(-q_f q_{-f} +p_f p_{-f})-\Delta(k_i)-1}
}
\nonumber\\
&
e^{-\oh \alpha' \sum_{c>0} R^2(\nu_c) \hat k_c \hat k_{-c}}\, 
(2\pi)^{d_K} \delta^{d_K}(q_f+k_f+p_f)\,
\langle-\lambda| e^{i k_c x^c_0} |\mu \rangle
\end{align}

\subsection{The stringy instantonic form factor}
It is now immediate to compute the form factor of $D(-1)$ seen from
$D3$.
Using the notation of (\cite{Billo:2002hm} we get
\begin{align}
\label{corr5} 
A^I_\mu(p;\bar w, w) = i\, (T^I)^{v}_{~u}\,p^\nu
\, \bar\eta^c_{\nu\mu}
\left(w_{\dot\alpha}^{~u}\,(\tau_c)^{\dot\alpha}_{~\dot\beta}\,
\bar w^{\dot \beta}_{~v}\right) 
\,e^{-i p\cdot x_0}
\,e^{-2 \ln2\, \alpha' p_\nu p^\nu}
\end{align}
where $\mu=0,1,2,3$ and we have used $\Delta_{\nu=\oh}(x)=-2\ln(1+\sqrt{x})$ so that
$R^2(\oh)=4\ln 2$.  We have also used  $F_0=0$.
It is interesting to notice that only for $\nu=\oh$ the $\Theta_{vect}$
term vanishes. 
It then follows that an instanton cannot shrink to zero and that its
profile is 
\begin{eqnarray}
\label{gf1} 
A^I_\mu(x) 
&=& 
\int {d^4 p\over (2\pi)^2} \,
A^I_\mu(p; \bar w, w) \,{1\over p^2}\,e^{i p\cdot x} 
\nonumber\\ 
&=& (T^I)^{v}_{~u} \, \left(w_{\dot\alpha
}^{~u}\,(\tau_c)^{\dot\alpha}_{~\dot\beta}\, \bar w^{\dot \beta
}_{~v} \right)\,\bar\eta^c_{\nu\mu} \, 
\left\{
\begin{array}{c c}
-2 {(x-x_0)^\nu\over(x-x_0)^4} &~~~~  (x-x_0)^2 >> \alpha' R^2(\oh)
\\
-\frac{2}{3}\frac{(x-x_0)^\nu }{\alpha'^2 R^4 } &~~~~  (x-x_0)^2 << \alpha' R^2(\oh)
\end{array}
\right.
\end{eqnarray}
Since instantonic amplitudes on non compact space are dominated by the
infrared behavior of the instanton it is then clear that such
amplitudes essentially are the same as without the form factor. 
On the contrary the form factor can give relevant contributions when
the space is compact and the size is not too large.

\noindent {\large {\bf Acknowledgments}}
\vskip 0.2cm
\noindent 

We would like to thank M. Bill\`o, P. Di Vecchia, A. Lerda,  R. Marotta and
F. Pezzella for several discussions. The author thanks
the Niels Bohr Institute for hospitality during different stages of
this work.

\appendix
\section{Conventions.}
\label{conventions}
We define:
\begin{itemize}
\item 
WS metric signature: $\eta_{\alpha \beta }=(-,+)$; $\epsilon^{ 0 1}=-1$ 
\\ 
Space-time metric signature: $G_{\mu \nu}=(-,+,\dots, +)$;
\item Indices:
Compact $i,j,\dots= 1,\dots d$;
non compact $\mu,\nu,\dots= 0,d+1\dots D$;
general $M,N,\dots=0,\dots D$;
\item 
WS coordinates: $\xi_\pm= \tau \pm \sigma$ and $\tau_E=i \tau$
\begin{equation}
z=e^{\tau_E+i\sigma}\,\,\,\,\,\,\,\,,\,\,\,\,\,\,\,\,
\bar z=e^{\tau_E-i\sigma}~~,
\end{equation}
the variables $z$ and $\bar z$ are defined respectively
in the upper-half and in the lower-half complex plane since 
$\sigma\in[0,\pi]$
\item
Given a couple of free bosons $\un X^{1,2}\equiv X^{\un1, \un 2}$
(which we identify with flat coordinates)
we make the linear combinations
\begin{equation}
\un Z=\un X=\frac{1}{\sqrt{2}}\left( \un X^1+ i \un X^2 \right)
\,\,\,\,\,\,
\bar{ \un Z}=\bar{ \un X} = \un Z^\dagger
\end{equation}
Notice that in the main text we do not write the underline explicitly.
\item
We define the following functions which are used in the definitions of
``propagators''   for $0<\epsilon<1$ 
\begin{eqnarray*}
\hat g_\epsilon(z)
&=&
 g_\epsilon(z)
=
-\sum_{n=1}^{\infty}
\frac{1}{n-\epsilon} z^{n-\epsilon}
~~~~~
|z|<1, |arg(z)|<\pi
\nonumber\\
\hat g_{-\epsilon}(z)
&=&
 g_{1-\epsilon}(z)
\end{eqnarray*}
\item Background matrices:\\
\begin{eqnarray}
E&=& \parallel E_{i j} \parallel = G+B
\nonumber\\
{\cal E} &=& \parallel \cE_{i j} \parallel=E^T + 2\pi \alpha' q_0 F
= G +\cF
\end{eqnarray}
and
\begin{eqnarray}
\hat F &=& 2\pi \alpha' q_0 F
\nonumber\\
\cF  &=&  2\pi \alpha' q_0 F-B = \hat F-B
\end{eqnarray}
For the dipole strings we can define
\begin{equation}
{\cal E}^{-1} = {\cal G}^{-1} -\Theta
\end{equation}
from which we deduce that
\begin{eqnarray}
&&{\cal E}{\cal G}^{-1} {\cal E}^T
={\cal E}^T{\cal G}^{-1} {\cal  E}
=G
\nonumber\\
&&
\Theta= \frac{1}{2}\left({\cal E}^{-T}-{\cal E}^{-1} \right)
=- {\cal E}^{-1} {\cal B} {\cal E}^{-T}
\end{eqnarray}

\end{itemize}

\section{Details on the dicharged string quantization.}
\label{appDetailsQuant}
Using the product definition (\ref{AntiHermProd}) and the
eigenfunctions
(\ref{Eigenfunctions})
we find the following basic products 
\begin{align}
\langle \Psi_{n\, a}, \Psi_{m\,b}\rangle
&=
-i \pi ~\delta_{a,b} ~\delta_{m,n} ~(n+\nu_a)
&n+\nu_a,m+\nu_b\ne 0
\nonumber\\
\langle \Psi_{0\,e}, \Psi_{m\,a}\rangle
&=0
&\nu_e=0, m+\nu_a\ne0
\nonumber\\
\langle w, \Psi_{m\,a}\rangle
&=0
&m+\nu_a\ne 0
\nonumber\\
\langle \Psi_{0\,e}, \Psi_{0,f}\rangle
&=
-\pi^2~ v_e^\dagger~\cF_\pi~ v_f
&
\nu_e,\nu_f=0
\nonumber\\
\langle w, \Psi_{0\,e}\rangle
&=
\pi w^\dagger ~\cE_\pi^T~ v_e
&
\nu_e=0
\nonumber\\
\langle w, u\rangle
&=
-w^\dagger~\Delta \cF~ u
&~
\label{AppPsi-Psi}
\end{align}
where we have defined $\Delta \cF = \cF_\pi -\cF_0$ and $w,u$ are
arbitrary constant vectors.
The computation is almost straightforward when using the right tricks
therefore we show a piece of it:
\begin{align}
&\Psi_{n\, a}^\dagger(\pi,\tau) 
~\cFpi
~\Psi_{m\, b}(\pi,\tau)
=
\nonumber\\
&\qquad=
(-)^{m+n}
e^{i [(n+\nu_a)-(m+\nu_b)]\tau}
~v^\dagger_a ~ e^{i \pi \nu_a} \frac{\uno+ e^{-2 i \pi \nu_a} R_0^T}{2}
\cFpi
 \frac{\uno+  R_0 e^{-2 i \pi \nu_a} }{2} e^{-i \pi \nu_b}~v_b
\nonumber\\
&\qquad=
(-)^{m+n}
e^{i [(n+\nu_a)-(m+\nu_b)]\tau}
~v^\dagger_a ~ e^{i \pi \nu_a} \frac{\uno+ R_\pi^T}{2}
\cFpi
 \frac{\uno+  R_\pi }{2} e^{-i \pi \nu_b}~v_b
\nonumber\\
&\qquad=
(-)^{m+n}
\frac{1}{4}
e^{i [(n+\nu_a)-(m+\nu_b)]\tau}
~v^\dagger_a ~ e^{i \pi \nu_a} 
~G (R_\pi^{-1} - R_\pi)
 e^{-i \pi \nu_b}~v_b
\nonumber\\
&\qquad=
(-)^{m+n}
\frac{1}{4}
e^{i [(n+\nu_a)-(m+\nu_b)]\tau}
~
\left(
e^{-i \pi (\nu_a+\nu_b)}
~v^\dagger_a ~   G R_0^{-1} ~v_b
-
e^{+i \pi (\nu_a+\nu_b)}
~v^\dagger_a ~   G R_0 ~v_b
\right)
\end{align}
where we have used a certain number of times $R_\pi= R_0 R^{-1}$,
we have rewritten $\cFpi= G( \uno -R_\pi)(\uno+R_\pi)^{-1}$
and  used the definition of the eigenvalues $v_a$ in eq. (\ref{R-eigen})
and the orthogonality relation for $R$ in eq. (\ref{R-ortho}).
We can now define the quantities
\begin{align}
I_{a\,n}&=
\langle \Psi_{a\,n} , \Phi \rangle
=
-i \pi (n+\nu_a)~\phi^a_n
&
n+\nu_a\ne 0
\nonumber\\
I_{w=\cE^{-1}_\pi G v_e}&=
\langle \cE^{-1}_\pi G v_e , \Phi \rangle
=
\pi \phi^e_0
& 
\nu_e=0
\label{AppdefI0}
\end{align}
from which we can easily extract the coefficients $\phi^a_n$. 
The analogous expressions for  $\phi^i$ give
\begin{align}
I_{e}&=
\langle \Psi_{0\,e} , \Phi \rangle
=
-\pi~ v_e^\dagger \left( \cE_\pi~\phi 
                     + \pi ~\cF_\pi~v_f~ \phi^f_0\right)
& 
\nu_e=0
\nonumber\\
I_{w=\cE^{-1}_\pi G v_c}&=
\langle \cE^{-1}_\pi ~G ~v_c , \Phi \rangle
=
-v_c^\dagger~G~\cE_\pi^{-T}~\Delta\cF~\phi
&~ 
\label{AppdefI}
\end{align}
from which  it
is possible in principle to recover the $\phi^i$ coefficients 
since the equations in the last two lines are
as many as the components of $\phi^i$ and are independent 
and are therefore sufficient but the result is rather not instructive.
In order to compute the commutation relation we resort therefore to
the trick of computing the commutation relations among the previous
quantities (\ref{defI0}, \ref{defI}) when we take $\Phi=X$ 
and then extract those among the $x^i$.

Using the canonical commutation relations (\ref{CanComRel}) 
the commutation relations among the previous defined
quantities (\ref{AppdefI0}, \ref{AppdefI}) when $\Phi=X$ are
\begin{align}
[I_{n\,a}~,~I_{m\,b}]
=&
2 \pi^2 \alpha' (n+\nu_a) \delta_{a,b} \delta_{m,n}
\nonumber\\
[I_e~,~I_f]
=&
-i 2\pi^3 \alpha'~v_e^\dagger~\cF_\pi~v_f^*
\nonumber\\
[I_e~,~I_w]
=&
-i 2\pi^2 \alpha'~v_e^\dagger~\cE_\pi~w^*
\nonumber\\
[I_w~,~I_u]
=&
-i 2\pi \alpha'~w^\dagger~\Delta\cF~u^*
\end{align}
from which we can deduce the commutation relations for the string operators
written in the main text.

\section{Analytic properties of the twisted propagator.}
From the normal ordering it turns out natural to
define the following function which are used in the definitions of ``propagators'' for $0<\epsilon<1$ 
\begin{eqnarray}
g_\epsilon(z)
&=&
-\sum_{n=1}^{\infty}
\frac{1}{n-\epsilon} z^{n-\epsilon}
~~~~~
|z|<1, -\pi< arg(z)< \pi
\end{eqnarray}
We will also use the following more compact notation for $-1<\epsilon<1$
\begin{equation}
\hat g_{\epsilon}(z)
=\left\{\begin{array}{cc}
g_\epsilon(z) & 0<\epsilon<1 \\
g_{1+\epsilon}(z) & -1<\epsilon<0
\end{array}
\right.
\end{equation}
The reason of the range of $arg(z)$ is that we put a cut along the
negative real axis.
The previous function can be defined on all sheets as
\begin{equation}
g_{\epsilon,s}(z)
=
-\sum_{n=1}^{\infty}
\frac{1}{n-\epsilon} z^{n-\epsilon}
~~~~~
|z|<1, -\pi+2\pi s< arg(z)< \pi+ 2\pi s
\end{equation}
Its analytic continuation to the whole complex plane but the negative
real axis and $ x\ge 1$ part of the real one 
can be given by the following integral
representation which trivially reduces to the previous series in the
$|z|<1$ domain
\begin{eqnarray}
g_{\epsilon,s}(z)
&=&
- \int_0^{|z|} d x ~ \frac{x^{-\epsilon} e^{i (1-\epsilon) \phi}}{1- e^{i \phi} x} 
~~~~
z= |z| e^{i \phi}\in \C -\{x<0, x\ge 1\}
\\
&=&
-\int_\gamma d t ~ \frac{t^{-\epsilon}}{1-t} 
\end{eqnarray}
where the path $\gamma$ is the straight line from $t=0$ to $t=z$.
The cut along $x\ge 1$ is because of the logarithmic singularity at
$z=1$ which is due to the integrand pole at $t=1$.

Given the previous definition it is then easy to see that
\begin{eqnarray}
g_{\epsilon,s}(z)
= 
\int_0^{+\infty} d y \frac{y ^{-(1-\epsilon)} e^{i(1-\epsilon)\phi} }
{  e^{i \phi} -y}
+ g_{1-\epsilon,-s}\left(\frac{1}{z}\right)
=
C_{\epsilon,s}(arg~z)+ g_{1-\epsilon,-s}\left(\frac{1}{z}\right)
\label{g-an-cont}
\end{eqnarray}
As it is usual in complex analysis we consider the auxiliary integral
$\int d w \frac{w ^{-\epsilon } } {  e^{i \phi} +w}$ and then we can 
easily compute the constant entering the previous relation to be
\begin{equation}  
C_{\epsilon,s}(arg~z)=
\int_0^{+\infty} d y \frac{y ^{-\epsilon} }{  e^{i \phi} -y}
=
\left\{ \begin{array} {cc}
\frac{\pi e^{-i \pi \epsilon}}{sin \pi\epsilon} e^{-i 2\pi\epsilon s}&
2\pi s<\phi<\pi+2\pi s
\\
\frac{\pi e^{+i \pi \epsilon}}{sin \pi\epsilon} e^{-i 2\pi\epsilon s} & 
-\pi+2\pi s<\phi<2\pi s
\end{array}
\right.
\label{Cepss}
\end{equation}
The previous expression satisfies the following properties
\begin{equation}
C_{1-\epsilon,-s}(arg~1/z)= - C_{\epsilon,s}(arg~z),
~~~~
[C_{\epsilon,s}(arg~z)]^*=C_{\epsilon,-s}(arg~z^*),
\end{equation}
which can be derived as consistency conditions from
eq. (\ref{g-an-cont}) and verified directly on the expression (\ref{Cepss}).
It is then immediate to write the corresponding property for the $\hat
g$ function:
\begin{eqnarray}
\hat g_{\epsilon,s}(z)
=
\hat C_{\epsilon,s}(arg~z)+ \hat g_{-\epsilon,-s}\left(\frac{1}{z}\right)
,~~~~
\hat C_{\epsilon,s}(arg~z)=
\left\{ \begin{array} {cc}
\frac{\pi e^{-i \pi \epsilon}}{sin \pi\epsilon} e^{-i 2\pi\epsilon s}&
2\pi s<\phi<\pi+2\pi s
\\
\frac{\pi e^{+i \pi \epsilon}}{sin \pi\epsilon} e^{-i 2\pi\epsilon s} & 
-\pi+2\pi s<\phi<2\pi s
\end{array}
\right.
\label{ghat-an-cont}
\end{eqnarray}

It is also useful to compare the behavior of $g_\epsilon(z)$ with that of the
usual untwisted propagator $g_0(z)=\log(1-z)$, then we get for $\eta>0$
\begin{equation}
g_{\epsilon,s}(1-\eta)= 
\log \eta  
+\int_0^{1-\eta} d t~ \frac{1-t^{-\epsilon}}{1-t}
=
\log \eta
+\psi(1-\epsilon)-\psi(1)
+\epsilon \eta
+\frac{\epsilon(\epsilon+1)}{4} \eta^2
+ O(\eta^3)
\end{equation}
where we used the fact that $\psi= \frac{d \log \Gamma(z)}{d z}$ 
can be expressed as 
\begin{equation}
\psi(1+z)
=
-\gamma
+ \sum_{n=1}^\infty \frac{z}{n(n+z)} 
=
-\gamma
+\int_0^1 d t~ \frac{1-t^z}{1-t}
\label{psi_sum}
\end{equation}
with $\gamma$  the Euler-Mascheroni constant.
We record here a useful $\gamma$ property which we use in the computations 
\begin{equation}
\psi(1-z)= \psi(z) + \pi ~\cot \pi z
\end{equation}
so that taking $\epsilon=\nu_a$ we get
\begin{equation}
\Delta_a(1)-\Delta_{-a}(1)=  \pi ~\cot \pi \nu_a
\end{equation}

\section{Dipole open string conventions}
\label{app:Dipoleopen}

\subsection{Real formalism}

On the magnetic background  $\cF_0=\cF_\pi$ 
the open dipole string field expansion is given by
\begin{eqnarray}
X^i(z,\bar z) &=&
\frac{1}{2}\left(X_L^i(z)+ X_R^i(\bar z) \right)
\label{X-open-dipole}
\end{eqnarray}
where $z=e^{\tau_E +i \sigma}\in \Hp$ ($0\le \sigma \le \pi$) and
\begin{eqnarray}
X_L^i(z)
&=&
(G^{-1} {\cal E})^i_j
\hat X_{L (0)}^j(z)
+y_0^i
\nonumber\\
X_R^i(\bar z)
&=&
(G^{-1} {\cal E}^T)^i_j
\hat X_{R (0)}^j(\bar z)
-y_0^i
\label{hatXy0Def}
\end{eqnarray}
where we have defined the following quantities as in eq. (\ref{def-cE}) 
\begin{eqnarray}
{\cF}_{i j} 
&=& 
-B_{i j} + \hF_{i j}
\label{calB}
\\
{\cal{E}}_{i j} 
&=& 
G_{i j} +{\cF}_{i j} 
= G_{i j}
- B_{i j} + \hat F_{i j}
\label{calE}
\end{eqnarray}
and the open string metric given by
\begin{eqnarray}
{\cal{G}}_{i j} =  G_{i j} - {\cF}_{i k} G^{k h} {\cF}_{h j} =
{\cal{E}}^{T}_{i k} G^{k h} {\cal{E}}_{h j}
\label{openme2}
\end{eqnarray}
along with the non commutativity parameter $\Theta$ as
\begin{equation}
(\cE^{-1})^{i j}= (\cG^{-1})^{i j} - \Theta^{i j}~.
\end{equation}
Moreover we have also defined the fields $\hat X_{L (0)}^i$ and $\hat X_{R (0)}^i$
which have the usual field expansion
\begin{eqnarray}
\hat X_{L (0)}^i(z)
&=&
\hat x^i 
-2\alpha'  \hat p^i ~i \ln(z)
+ i \sqrt{2\alpha'} \sum_{n\ne 0} \frac{sgn(n)}{\sqrt{|n|}} \hat a_n^i
z^{-n}
\nonumber\\
\hat X_{R(0)}^i(\bar z)
&=&
\hat x^i 
-2\alpha'  \hat p^i ~i \ln(\bar z)
+ i \sqrt{2\alpha'} \sum_{n\ne 0} \frac{sgn(n)}{\sqrt{|n|}} \hat a_n^i
{\bar z}^{-n}
\label{hatXExp}
\end{eqnarray}
but have a different set of commutation relations since  the metric
$G$ is replaced by the open string metric $\cG$ and the zero modes
have a non trivial commutation relation, explicitly
\begin{eqnarray}
[\hat x^i, \hat x^j]= i ~2\pi\alpha' ~\Theta^{i j}
~~~~
[\hat x^i, \hat p^j]= i {\cal G}^{ i j}
~~~~
[\hat a_m^i,\hat a_n^j] = {\cal G}^{i j} ~sgn(m) ~\delta_{n+m,0}
~~~.
\end{eqnarray}
The previous expansion (\ref{hatXExp}) looks as the usual one because 
we have used $\hat x^i$ with non vanishing commutator 
and not $\hat x_0^i$ defined as 
$\hat x^i= \hat x_0^i - \pi\alpha' \Theta^{i l} {\cal  G}_{l m} \hat p^m$ 
with the usual commutator
\begin{equation}
[\hat x_0^i, \hat x_0^j]=0
\end{equation}
Finally $y_0^i= \sqa G^{i j} \theta_j$ are constants 
and proportional to the Wilson lines $\theta$ on the brane at $\sigma=0$
(\cite{Green:1991et},\cite{Frau:1997mq}, \cite{Pesando:2003ww} and
\cite{Pesando:2009tt} ).
Notice the $y_0^i$ do  not enter the expansion of $X(z, \bar z)$  and
therefore they do enter the open string vertexes but they do enter the
closed string vertexes in the open string formalism where they are
necessary to reproduce from the open string side the phases in the
boundary state due to Wilson lines or positions.

On this background the dipole string Hamiltonian  is then given by
\begin{eqnarray}
H_o-1=L_0 
&=& 
\alpha' G^{\mu\nu} k_\mu k_\nu
+\alpha'  \hat p^i {\cal{G}}_{i j} \hat p^j 
+  \sum_{n=1}^{\infty} n 
{{G}}_{\mu \nu} \alpha^{\dagger \mu}_{n} \alpha^{\nu}_{n} 
+n {\cal{G}}_{i j} \hat a^{\dagger i}_{n} \hat a^{j}_{n} 
\label{L061}
\end{eqnarray}

In the non compact case the vacuum is defined 
\begin{equation}
a^i_n | 0 \rangle = p^i | 0 \rangle  =0
,~~~~ n>0 
\end{equation}
so that the basic OPE reads
\begin{align}
X^{(-)i}_L(z) X^{(+)j}_L(w)
&=
-2\alpha' G^{i j} \log(z-w) 
+
:X^{(-)i}_L(z) X^{(+)j}_L(w):
\nonumber\\
&=
-2\alpha' G^{i j} \left(\log(z)+ g_0\left(\frac{w}{z}\right)\right)
+\dots 
\end{align}

\subsection{Complex formalism in two dimensions.}
We consider $\R^2$ on which we write the background matrices in flat
coordinates as
\begin{equation}
\un \cE
=
\un G -\un B + \un \hF
=
\left(\begin{array}{c c}
1 & \un \ff \\ 
-\un \ff & 1 
\end{array}\right),
~~~~
\un \ff = \un {\hat \ff} -\un \bb
\end{equation}
from which we derive the open string background
\begin{equation}
\un \cG=\frac{1}{1+ \un \ff^2}
\left(\begin{array}{c c}
1 & 0 \\ 
0 & 1 
\end{array}\right)
=\cos^2 \gamma ~\uno_2
,~~~~
\un \Theta = 
\frac{1}{1+ \un \ff^2}
\left(\begin{array}{c c}
0 & \un \ff \\ 
-\un \ff & 0 
\end{array}\right)
=\cos \gamma \sin \gamma ~\epsilon_2
\end{equation}
where we have defined the angle $\gamma$ as
\begin{equation}
e^{i \gamma}= \frac{1+ i \un \ff}{\sqrt{1+ \un \ff^2}}
\Rightarrow
\un \ff = \tan \gamma
,~~~~
-\frac{\pi}{2}<\gamma<\frac{\pi}{2}
\end{equation}

If we define the complex fields
\begin{equation}
\un Z(z,\bar z)
=
\frac{1}{\sqrt{2}}\left( \un X^1 (z,\bar z) +i \un X^2 (z,\bar z) \right)
=
\frac{1}{2}( \un Z_L(z) + \un Z_R(\bar z))
\end{equation}
the boundary conditions become 
\begin{equation}
\un Z ' + i \un \ff \un {\dot  Z}  |_{\sigma=0} =0
,~~~~
\un Z ' + i \un \ff \un {\dot  Z}  |_{\sigma=\pi} =0
.
\label{A-Z-bou-cond}
\end{equation}
or in the worldsheet Wick rotated version
\begin{eqnarray*}
e^{+i\gamma} \partial \un Z |_x &=& 
e^{-i\gamma} \bar\partial \un Z |_{x} 
\spazio
x\in \R^+
\\
e^{+i\gamma} \partial \un Z |_y &=& 
e^{-i\gamma} \bar\partial \un Z |_y 
\spazio
y= |y| e^{i\pi}\in \R^-
\end{eqnarray*}
The fields expansions for $\un Z$ read
\begin{eqnarray}
\un Z_L 
&=&
(1- i \un \ff) \hat{ \un Z}_{L (0)}+ w_0
= e^{-i \gamma}  { \un Z}_{L (0)}+ w_0
\nonumber\\
&=&
e^{-i \gamma} \Bigg(
\un z_0 
-2\alpha'  \un p ~i \ln(z)
+ i \sqrt{2\alpha'} \sum_{n=1}^\infty 
+\frac{\bar { \un a}_n }{\sqrt{n}}  z^{-n}
-\frac{\un a_n^\dagger }{\sqrt{n}}  z^{n}
\Bigg)
+ w_0
\nonumber\\
\un Z_R
&=&
(1+ i \un \ff) \hat{ \un Z}_{R (0)} - w_0
= e^{+i \gamma}  { \un Z}_{R (0)} - w_0
\nonumber\\
&=&
e^{+i \gamma} \left(
\un z_0 
-2\alpha'  \un p ~i \ln(\bar z)
+i \sqrt{2\alpha'} \sum_{n=1}^\infty
+\frac{\bar { \un a}_n }{\sqrt{n}}  \bar z^{-n}
-\frac{\un a_n^\dagger }{\sqrt{n}}  \bar z^{n}
\right)
- w_0
\end{eqnarray}
where we have introduced 
${ \un Z}_{L (0)}= \sqrt{1+\un \ff^2 }\hat{ \un Z}_{L (0)}$
and for their complex conjugates $\un {\bar
  Z}=\frac{1}{\sqrt{2}}\left(\un X^1- i \un X^2\right)$  
\begin{eqnarray}
\un {\bar Z}_L 
&=&
(1+ i \un \ff) \hat{ \bar{ \un Z}}_{L (0)}+ \bar w_0
= e^{+i \gamma}  { \un {\bar Z}}_{L (0)}+ \bar w_0
\nonumber\\
&=&
e^{+i \gamma} \left(
\un {\bar z}_0 
-2\alpha'  \un {\bar p} ~i \ln(z)
+ i \sqrt{2\alpha'} \sum_{n\ne 0} 
-\frac{\bar { \un a}_n^\dagger }{\sqrt{n}}  z^{n}
+\frac{\un a_n }{\sqrt{n}}  z^{-n}
\right)
+ \bar w_0
\nonumber\\
\un {\bar Z}_R
&=&
(1- i \un \ff) \hat{ \un {\bar Z}}_{R (0)} - \bar w_0
= e^{-i \gamma}  { \un {\bar Z}}_{R (0)} - \bar w_0
\nonumber\\
&=&
e^{-i \gamma} \left(
\un {\bar z}_0 
-2\alpha'  \un {\bar p} ~i \ln(\bar z)
+ i \sqrt{2\alpha'} \sum_{n\ne 0} 
-\frac{\bar { \un a}_n^\dagger }{\sqrt{n}}  \bar z^{n}
+\frac{\un a_n }{\sqrt{n}}  \bar z^{-n}
\right)
- \bar w_0
\end{eqnarray}
with commutation relations
\begin{eqnarray}
[\un z_0, \bar{\un z}_0 ]&=& 2\pi \alpha' \un \ff
\nonumber\\
{}[\un z_0, \bar{\un p}_0 ]&=& i
\nonumber\\
{}[\un a_n ,\bar{\un a}_m] &=& \delta_{m+n,0}~ sgn(m)
\end{eqnarray}

When we define as usual ($n>0$)
\begin{eqnarray*}
\un \alpha_n =\sqrt{n} \un a_n
\spazio
\un\alpha_{-n} =\sqrt{n}\un a_n^\dagger
\spazio  
\un \alpha_0 = \sqrt{2\alpha'}\un p
\end{eqnarray*}
and similarly for the barred quantities, we can write the
energy-momentum tensor as
\begin{eqnarray*}
T(z) &=&
-\frac{2}{\alpha'} \partial \un Z \partial \bar {\un Z} 
=
-\frac{1}{2\alpha'} \partial \un Z_L \partial \bar {\un Z}_L 
= \sum_k \frac{L_{k  }}{z^{k+2}}
\end{eqnarray*}
with
\begin{eqnarray}
L_{k } &=& 
\sum_{m=1}^\infty \un{\bar \alpha}_{m }^\dagger \;\un{\bar \alpha}_{m +k } 
+ \sum_{m=0}^\infty \un\alpha_{m }^\dagger\;\un\alpha_{m + k } 
+ \sum_{m=1}^{k} \un{\bar \alpha}_{m }\; \un\alpha_{k - m }
\spazio k>0
\nonumber\\
L_ {k } &=& 
\sum_{m=1}^\infty \un{\bar \alpha}_{m + |k| }^\dagger
                \;\un{\bar \alpha}_{m } 
+ \sum_{m=0}^\infty \un\alpha_{m + |k| }^\dagger
                   \;\un\alpha_{m } 
+ \sum_{m=1}^{|k|} \un{\bar \alpha}_{m }^\dagger\;
                   \un\alpha_{|k| - m }^\dagger
\spazio
k<0
\nonumber\\
L_{0 }& =& 
\alpha_0 \un{\bar\alpha}_0
+\sum_{m=1}^\infty \un{\bar \alpha}_{m }^\dagger \;\un{\bar \alpha}_{m } 
+ \sum_{m=1}^\infty \un\alpha_{m }^\dagger \;\un\alpha_{m } 
\nonumber\\
& =&  \alpha' \un { p}  \un {\bar p}
+ \sum_{n=1}^\infty
n\left(
{\un a}^\dagger_{n} \un a_n +
\bar {\un a}^\dagger_{n} \bar {\un a}_n
\right)
\end{eqnarray}

The tachyonic vertex is then
\begin{eqnarray}
V_T(x, k)
&=&
: e^{i k_i X^i(x,x)}: ~\Lambda
\nonumber\\
&=&
: e^{i k_i \hat X^i_{L(0)}(x)}: ~\Lambda
\nonumber\\
&=&
: e^{i  ( 
\un k e^{i \gamma} \hat {\bar{\un Z}}_{L(0)}(x)
+
\bar{ \un k} e^{-i \gamma} \hat {\un Z}_{L(0)}(x)
)
}: ~\Lambda
\nonumber\\
&=&
: e^{i \cos \gamma ( 
\un k e^{i \gamma} \bar{\un Z}_{L(0)}(x)
+
\bar{ \un k} e^{-i \gamma} {\un Z}_{L(0)}(x)
)
}: ~\Lambda
\end{eqnarray}
where the normal ordering for the zero modes is the trivial one, i.e.
$:e^{i k_i \hat x^i}: = :e^{i k_i \hat x^i}$ as suggested by the
hermiticity  of the vertex.
The $\Lambda$ are the hermitian Chan-Paton matrices.
The previous tachyonic vertex has  conformal dimension
\begin{equation}
\Delta(V_T)= \alpha' \cG^{i j}k_i k_j = 2 \alpha' \cos^2\gamma~\un k
~\bar{\un k}
\end{equation}
and OPE
\begin{eqnarray}
V_T(x_1,k ) V_T(x_2,l)
&=&
e^{ \frac{1}{2} ~2 \pi \alpha'~\cos (2 \gamma) 
~( \un k \bar{\un l}   - \un l \bar{\un k}  ) }
(x_1 -x_2)^{2\alpha' \cos^2 \gamma~( \un k \bar{\un l}   - \un l \bar{\un k}  ) }
: V_T~V_T:  
\nonumber\\
\end{eqnarray}
which must be reproduced by the corresponding emission vertex from the
dicharged string. 

\section{Dicharged string on $\R^2$ in complex formalism.}
For dicharged strings on $\R^2$ it is natural to work in complex formalism
 which we use in section \ref{sect:tach-vert-R2}.
Here we state our normalizations and conventions in this formalism.
Using flat coordinates  the boundary conditions are
\begin{eqnarray}
\un Z'+i \un \ff_{(0)} \un{ \dot Z} |_{\sigma=0}=0
,~~~~
\un Z'+i \un \ff_{(\pi)} \dot {\un Z}|_{\sigma=\pi}=0
\label{app-Z-bou-cond}
\end{eqnarray}
or in the worldsheet Wick rotated version
\begin{eqnarray}
e^{i\gamma_0} \partial \un Z |_x &=& 
e^{-i\gamma_0} \bar\partial \un Z |_{x} 
\spazio
x\in \R^+
\\
e^{+i\gamma_\pi} \partial \un Z |_y &=& 
e^{-i\gamma_\pi} \bar \partial \un Z |_y 
\spazio
y= |y| e^{i\pi}\in \R^-
\end{eqnarray}
where we have defined as for the dipole string
\begin{eqnarray}
e^{i\gamma_0}= \frac{1 + i \un \ff_0 }{\sqrt{1+\un \ff_0^2}}
&\rightarrow&
\un \ff_0 = \tan \gamma_0
\nonumber\\
e^{i\gamma_\pi}= \frac{1 + i \un \ff_\pi }{\sqrt{1+\un \ff_\pi^2}}
&\rightarrow&
\un \ff_\pi = \tan \gamma_\pi
\end{eqnarray}
It is then immediate to compute the 
\begin{align}
\un R_{0,\pi} = \left(\begin{array}{c c}
\cos 2\gamma_{0,\pi} & -\sin 2\gamma_{0,\pi}\\
\sin 2\gamma_{0,\pi} & \cos 2 \gamma_{0,\pi}
\end{array}
\right)
~\Rightarrow~
\un R = \left(\begin{array}{c c}
\cos 2(\gamma_0-\gamma_\pi) & -\sin 2(\gamma_0-\gamma_\pi)\\
\sin 2(\gamma_0-\gamma_\pi) & \cos 2 (\gamma_0-\gamma_\pi)
\end{array}
\right)
\end{align}
so that we 
\begin{eqnarray}
\nonumber\\
1> \epsilon =\nu &=& \frac{|\gamma_0 - \gamma_\pi|}{\pi}>0
\end{eqnarray}
For the complex string fields
\begin{equation}
\un Z(z,\bar z)= \frac{1}{2}\left( \un Z_L(z)+\un Z_R(\bar z)\right)
,~~~~
\bar{\un Z}(z,\bar z)= 
\frac{1}{2}\left( \bar{\un Z}_L(z)+\bar{\un Z}_R(\bar z)\right)
\end{equation}
we have the following expansions\footnote{
Notice the redefinitions 
$e^{-i\gamma_0}  \bar {\un a}_{n+1-\epsilon}^{(usual)}=\bar {\un a}_{n+1-\epsilon}$
and
$e^{-i\gamma_0}  \un a_{n+\epsilon}^{(usual)\dagger}=\un a_{n+\epsilon}^\dagger$
for $n\ge 0$ as in \cite{Bertolini:2005qh}. 
}
\begin{eqnarray}
\un Z_L(z)
&=&\un \zlz
+i\sqrt{2\alpha'} 
\sum_{n=0}^{\infty}
\left[
\frac{ \bar {\un a}_{n+1-\epsilon} }{\sqrt{n+1-\epsilon}} z^{-(n+1-\epsilon)}
-
\frac{ \un a_{n+\epsilon}^\dagger}{\sqrt{n+\epsilon}} z^{+(n+\epsilon)}
\right]
\nonumber\\
\un Z_R(\bar z)
&=&
\un \zrz
+i\sqrt{2\alpha'} e^{+2 i\gamma_0} 
\sum_{n=0}^{\infty}
\left[ 
\frac{\bar{\un a}_{n+1-\epsilon} }{\sqrt{n+1-\epsilon}} {\bar z}^{-(n+1-\epsilon)}
- 
\frac{ \un a_{n+\epsilon}^\dagger}{\sqrt{n+\epsilon}} {\bar z}^{+(n+\epsilon)}
\right]
\end{eqnarray}
and
\begin{eqnarray}
\bar{\un Z}_L(z)
&=&\un{\bzlz}
+i\sqrt{2\alpha'} 
\sum_{n=0}^{\infty}
\left[
-\frac{ \bar{\un a}_{n+1-\epsilon}^\dagger}{\sqrt{n+1-\epsilon}} z^{+(n+1-\epsilon)}
+
\frac{ \un a_{n+\epsilon}}{\sqrt{n+\epsilon}} z^{-(n+\epsilon)}
\right]
\nonumber\\
\bar Z_R(\bar z)
&=& \un{\bzrz}
+i\sqrt{2\alpha'} e^{-2i\gamma_0} 
\sum_{n=0}^{\infty}
\left[ 
-\frac{\bar{\un a}_{n+1-\epsilon}^\dagger}{\sqrt{n+1-\epsilon}} {\bar z}^{+(n+1-\epsilon)}
+ 
\frac{\un a_{m+\epsilon}}{\sqrt{m+\epsilon}} {\bar z}^{-(m+\epsilon)}
\right]
\end{eqnarray}
with non vanishing commutation relations:
\begin{eqnarray}
[\un{\bar a}_{n+1-\epsilon} , \un{\bar a}_{m+1-\epsilon}^\dagger ]
&=&
\delta_{n,m} 
\spazio
n,m \ge 0
\nonumber\\
{}[\un{a}_{n+\epsilon} , \un{a}_{m+\epsilon}^\dagger ] 
&=&
\delta_{n,m} 
\spazio
n,m\ge 0
\nonumber\\
{}[\un z_0, \bar{\un z}_0]
&=&
\frac{2 \pi \alpha'}{ \un \ff_{(0)} -\un \ff_{(\pi)}}
\end{eqnarray}

The vacuum is the usual vacuum for the non zero modes  
and it is defined as
\begin{equation}
\un a_{n+\epsilon} | 0_\epsilon \rangle = 
\bar{\un a}_{n+1-\epsilon} |0_\epsilon \rangle=0
~~~~~n\ge 0
\end{equation}
while for zero modes it is necessary to distinguish between the non
compact case and the compact one.
For the non compact case we define it as
\begin{equation}
x^1 |0_\epsilon\rangle=0 ~~\rightarrow~~
|k,\epsilon \rangle = 
e^{-i {k}( {\un \ff_{0} - \un \ff_{\pi}})  x^2} |0_\epsilon\rangle=0
\end{equation}
which corresponds choosing $p=x^1/(2\pi\alpha')$  and 
$x= -(\un \ff_{(0)} - \un  \ff_{(\pi)}) x^2$, as noticed in the main
text the choice of $x^1$ ad destructor is arbitrary and we could have
chosen any linear combination of $x^1$ and $x^2$ as well.

The compact case is more subtle and it is discussed in a separate
paper \cite{Me}.

The left-right split of zero modes can be parametrized as
\begin{equation}
\un \zlz = (1+\alpha) \un z_0 + \beta \un{\bar z}_0+ \un w_0,
~~~~
\un \zrz = (1-\alpha) \un z_0 - \beta \un{\bar z}_0 - \un w_0,
\label{z0split}
\end{equation}
and because of the constraints from the closed OPE the parameters can
be fixed to $\alpha=\ff_0$  and $\beta=0$.

Energy-momentum tensor is given by
\begin{eqnarray}
T_\epsilon(z) =
-\frac{2}{\alpha'} :\partial \un Z \partial \bar{\un  Z}: 
- \frac{\epsilon(\epsilon-1)}{2 z^2}
=
-\frac{1}{2\alpha'} :\partial \un Z_L \partial \bar{\un  Z}_L:
-\frac{\epsilon(\epsilon-1)}{2 z^2}
= \sum_k \frac{L_{k  }}{z^{k+2}}
\end{eqnarray}
with
\begin{eqnarray}
L_{k (\epsilon)} &=& 
\sum_{m=1}^\infty \un{\bar \alpha}_{m - \epsilon}^\dagger 
\;\un{\bar \alpha}_{m +k - \epsilon} 
+ \sum_{m=0}^\infty \un{\alpha}_{m + \epsilon}^\dagger
\;\un{\alpha}_{m + k + \epsilon} 
+ \sum_{m=1}^{k} \un{\bar \alpha}_{m - \epsilon}\; \un{\alpha}_{k - m + \epsilon}
\spazio k>0
\nonumber\\
L_ {k (\epsilon)} &=& 
\sum_{m=1}^\infty \un{\bar \alpha}_{m + |k| - \epsilon}^\dagger
                \;\un{\bar \alpha}_{m - \epsilon} 
+ \sum_{m=0}^\infty \un{\alpha}_{m + |k| + \epsilon}^\dagger
                   \;\un{\alpha}_{m + \epsilon} 
+ \sum_{m=1}^{|k|} \un{\bar \alpha}_{m - \epsilon}^\dagger\;
                   \un{\alpha}_{|k| - m + \epsilon}^\dagger
\spazio
k<0
\nonumber\\
L_{0 (\epsilon)}& =& 
\sum_{m=1}^\infty \un{\bar \alpha}_{m - \epsilon}^\dagger
\;\un{\bar \alpha}_{m - \epsilon} 
+ \sum_{m=0}^\infty \un{\alpha}_{m + \epsilon}^\dagger
\;\un{\alpha}_{m + \epsilon} 
+ \frac{1}{2} \;\epsilon \;(1 - \epsilon)
\end{eqnarray}

\subsection{OPEs.}
Various useful contractions  are ($|z|>|w|$):
\begin{eqnarray}
\un Z_L^{(+,nzm)}(z)\, \bar{\un Z}_L^{(-,nzm)}(w)
&=&
-2\alpha'\, \hat g_\epsilon(w/z)
+:\dots:
%
\nonumber\\
\un Z_L^{(+,nzm)}(z)\, \bar{\un Z}_R^{(-,nzm)}(\bar w)
&=&
-2\alpha'\, e^{-2i \gamma_0} \hat g_\epsilon(\bar w/z)
+:\dots:
%
\nonumber\\
\un Z_R^{(+,nzm)}(\bar z)\, \bar{\un Z}_L^{(-,nzm)}(w)
&=&
-2\alpha'\, e^{2i \gamma_0} \hat g_\epsilon( w/ \bar z)
+:\dots:
%
\nonumber\\
\un Z_R^{(+,nzm)}(\bar z)\, \bar{\un Z}_R^{(-,nzm)}(\bar w)
&=&
-2\alpha'\,  \hat g_\epsilon(\bar w/\bar z)
+:\dots:
%
\end{eqnarray}
and
\begin{eqnarray}
\bar{\un Z}_L^{(+,nzm)}(z)\, {\un Z}_L^{(-,nzm)}(w)
&=&
-2\alpha'\, \hat g_{-\epsilon}(w/z)
+:\dots:
%
\nonumber\\
\bar{\un Z}_L^{(+,nzm)}(z)\, {\un Z}_R^{(-,nzm)}(\bar w)
&=&
-2\alpha'\, e^{2i \gamma_0} \hat g_{-\epsilon}(\bar w/z)
+:\dots:
%
\nonumber\\
\bar{\un Z}_R^{(+,nzm)}(\bar z)\, {\un Z}_L^{(-,nzm)}(w)
&=&
-2\alpha'\, e^{-2i \gamma_0} \hat g_{-\epsilon}( w/ \bar z)
+:\dots:
%
\nonumber\\
\bar{\un Z}_R^{(+,nzm)}(\bar z)\, {\un Z}_R^{(-,nzm)}(\bar w)
&=&
-2\alpha'\,  \hat g_{-\epsilon}(\bar w/\bar z)
+:\dots:
%
\end{eqnarray}


\section{Details on the vertexes construction on $\R^n$ and their consistency}
\label{AppDetailsVertexes}
In this appendix we give some details of the computations needed to
verify that the proposed vertexes and their generators do indeed
satisfy all the desired properties.
We first consider the problem in its full generality in one case by
computing the OPE of two generating vertexes for the emission of
dipole states from the $\sigma=0$ boundary of a dicharged string
then we exam all the other properties in a simpler setting, i.e. we
consider the tachyonic vertexes and $ker (\cF_\pi-\cF_0)=\emptyset$ only.

\subsection{OPE of the generating vertexes}
The OPE of two vertexes which describe the emission of a dipole
state from the dicharged string must have the same coefficients as the
OPE of the corresponding emission vertexes  from the dipole string
therefore
we start considering the OPE of two emission vertexes from a dipole
string.
We consider here the case of the emission from the $\sigma=0$ boundary
but we can obviously repeat the same procedure for the $\sigma=\pi$
vertexes.
 
More specifically we consider two generating vertexes for the emission
of dipole strings 
from a dipole string and we compute the product of two such vertexes to be
($x,y>0$)
\begin{align}
\cS_0(c,x)~ \cS_0(d,y)
&=
e^{i \pi \alpha' c_0^T \Theta d_0}
e^{-2\alpha' \sum_{k,l=0}^\infty c_k^T\,G^{-1}\, d_l~ 
\partial^k_x  \partial^l_y G_0(x,y)}
\cS_0(\{ d_k+\sum_{m=0}^k c_m \frac{(x-y)^{k-m}}{(k-m)!} \},y)
\end{align}
then we expect that the corresponding OPE for the emission vertexes
from a dicharged string to read
\begin{align}
\cS(c,x)~ \cS(d,y)
&=
e^{i \pi \alpha' c_0^T \Theta d_0}
e^{-2\alpha' \sum_{k,l=0}^\infty c_k^T\,G^{-1}\, d_l~ 
\partial^k_x  \partial^l_y G_0(x,y)}
\cS(\{ d_k+\sum_{m=0}^k c_m \frac{(x-y)^{k-m}}{(k-m)!} \},y)
\end{align}
since the OPE coefficients $
e^{i \pi \alpha' c_0^T \Theta d_0}
e^{-2\alpha' \sum_{k,l=0}^\infty c_k^T\,G^{-1}\, d_l~ 
\partial^k_x  \partial^l_y G_0(x,y)}$ must be the same when we map
$\cS_0$ into $\cS$.
Notice that in the previous expressions we have taken the parameters
$\{ c_k\}$ not to be constants but generic functions and that this
does not change the fact that $\cS(c,x)$ can be used as generating
function for  all vertexes. 
Moreover and more subtly we have expanded $\Xln(x)$ around $x=y$  but
this expansion is convergent only for a subset of the circle
$|x|=|y|=1$ because $\Xln$ contains both positive and negative frequencies.

Let us verify the previous equation explicitly.
The product of two generating vertexes $\cS$ reads
\begin{align}
\cS(c,x) \cS(d,y)
=&
\exp\{
-\alpha' \sum_{k,l=0}^\infty c_k^T 
{} \,\cE_0^{-T} G\, v_c ~ v_c^\dagger\, G \cE_0^{-1}\,
c_l ~
\partial^k_u|_{u=x}~\partial^k_v|_{v=x} ~\Delta_c(u/v)
\}
\nonumber\\
&
\exp\{
-\alpha' \sum_{k,l=0}^\infty d_k^T 
{} \,\cE_0^{-T} G\, v_c ~ v_c^\dagger\, G \cE_0^{-1}\,
d_l ~
\partial^k_u|_{u=y}~\partial^k_v|_{v=y} ~\Delta_c(u/v)
\}
\nonumber\\
&
\exp\{-2\alpha' \sum_{k,l=0}^\infty c_k^T 
{} \,\cE_0^{-T} G\, v_a ~ v_a^\dagger\, G \cE_0^{-1}\,
d_l ~
\partial^k_u|_{v=x}~\partial^k_v|_{u=y} ~\hat g_{-\nu_a}(u/v)\}
~\exp\{\oh c_0^T~[x,x^T]~d_0\}
\nonumber\\
&
:\exp\{\sum_k [ d_k+ \sum_{m=0}^k c_m \frac{(x-y)^{k-m}}{(k-m)!} ]
  \partial^k_y X(y,y)\}:
\end{align}
We then reexpress the last line using $\cS(\{d_k +\sum_{m=0}^k c_m
\frac{(x-y)^{k-m}}{(k-m)!}\},y)$ and we use the definition of
$\Delta_a$ 
so we find
\begin{align}
\cS(c,x)~ \cS(d,y)
&=
\exp\Big\{-\alpha' \sum_{k,l=0}^\infty c_k^T
{} \,\cE_0^{-T} G\, v_c ~ v_c^\dagger\, G \cE_0^{-1}\,
~d_l ~
\partial^k_u|_{v=x}~\partial^k_v|_{u=y} 
~[\Delta_c(v,u)-\Delta_{-c}(u,v)]
\Big\}
\nonumber\\
&
~\exp\{\oh c_0^T~[x,x^T]~d_0\}
\nonumber\\
&
e^{-2\alpha' \sum_{k,l=0}^\infty c_k^T~G^{-1} d_l~ \partial^k_x
  \partial^l_y G_0(x,y)}
\cS_0(\{ d_k+\sum_{m=0}^k c_m \frac{(x-y)^{k-m}}{(k-m)!} \},y)
\end{align}
Using the property of $\hat g_{-\nu_a}$ we see that we are left with
\begin{align}
&=
\exp\Bigg\{-\alpha'  c_0^T  
{} \,\cE_0^{-T} G\, v_c ~ v_c^\dagger\, G \cE_0^{-1}\,
~d_0 ~ 
  \left[\hat C_{-\nu_c}\left(arg \frac{x}{y}\right)
   -
   \hat C_{0}\left(arg \frac{x}{y}\right)\right]
+\oh c_0^T~[x,x^T]~d_0\Bigg\}
\nonumber\\
&e^{-2\alpha'   \sum_{k,l=0}^\infty c_k^T~G^{-1} d_l~ \partial^k_x
  \partial^l_y G_0(x,y)}
\cS(\{ d_k+\sum_{m=0}^k c_m \frac{(x-y)^{k-m}}{(k-m)!} \},y)
\end{align}
which reproduces the desired OPE 
when we check that the zero modes contribution is the non commutative phase 
$e^{i \pi \alpha' c_0^T \Theta d_0}$.

This is exactly what we do now by rewriting the exponential  as
\begin{align}
&
\frac{1}{\alpha'}
\left\{
-
\oh c_0^T [x_0, x_0^T] d_0
+
\alpha' \sum_c c_0^T \cE_0^T G v_c~ v_c^\dagger G\cE_0^{-T} d_0
  \left[\hat C_{-\nu_c}\left(arg \frac{x}{y}\right)
   -
   \hat C_{0}\left(arg \frac{x}{y}\right)\right]
\right\}
\nonumber\\
=&
-i\pi c_0^T\Big[
  \cE_\pi^{-1}G v_f ~v_f^\dagger\cF_\pi v_g ~v_g^\dagger G\cE_\pi^{-T}
+
  \cE_\pi^{-1}G v_c ~v_c^\dagger\cE_\pi w_{c_1} \frac{1}{f_{c_1}}
 ~ w_{c_1}^\dagger\cE_0^T v_d ~v_d^\dagger G\cE_\pi^{-T}
\nonumber\\
&+
  \cE_0^{-1}G v_c ~\frac{1+e^{-i 2\pi \nu_c}}{1-e^{-i 2\pi \nu_c}} 
  ~v_c^\dagger G\cE_0^{-T}
\Big] d_0
\nonumber\\
\end{align}
where we used the explicit expressions for $\hat C$ 
\begin{align}
=&
-i\pi c_0^T\Big[
  \cE_\pi^{-1}G v_f ~v_f^\dagger\cF_\pi v_g ~v_g^\dagger G\cE_\pi^{-T}
\nonumber\\
&+
 \frac{1}{1-e^{-i 2\pi \nu_d}} \left[
 -
  2  \cE_\pi^{-1}G v_c ~v_c^\dagger\cE_\pi w_{c_1} 
  ~w_{c_1}^\dagger G\cE_0^{-1}G v_d ~v_d^\dagger G\cE_\pi^{-T}
 +
  \cE_0^{-1}G v_d ~(1+e^{-i 2\pi \nu_d}) ~v_d^\dagger G\cE_0^{-T}
  \right]
\Big] d_0
\nonumber\\
\end{align}
where we used 
$\frac{1}{f_d} w_d^\dagger \cE_\pi^T v_c =
-\frac{2}{1-e^{-i 2\pi \nu_c}}
w_d^\dagger G \cE_0^{-1} G v_c
$ which can be derived from $R$ definition
\begin{align}
=&
-i\pi c_0^T\Big[
  \cE_\pi^{-1}G v_f ~v_f^\dagger\cF_\pi v_g ~v_g^\dagger G\cE_\pi^{-T}
+
 \frac{1}{1-e^{-i 2\pi \nu_d}} \times
\nonumber\\
& \times
\left[
 +
  2  \cE_\pi^{-1}G v_f ~v_f^\dagger\cE_\pi \cE_0^{-1}G v_d ~v_d^\dagger G\cE_\pi^{-T}
 -
  2  \cE_0^{-1}G v_d ~v_d^\dagger G\cE_\pi^{-T}
 +
  \cE_0^{-1}G (1+R) v_d   ~v_d^\dagger G\cE_0^{-T}
  \right]
\Big] d_0
\nonumber\\
\end{align}
where we have used the spectral representation for $G^{-1}$ to write
 $ w_{c_1} ~w_{c_1}^\dagger =G^{-1}- w_{f_1} ~w_{f_1}^\dagger$, then
we used the orthogonality of $G^{-1}\cE_\pi w_f\in ker\, \Delta R$ with
$v_c$ and finally we used again the spectral representation for
$G^{-1}$ to write $ v_{c} ~v_{c}^\dagger =G^{-1}- v_{f} ~v_{f}^\dagger$
\begin{align}
=&
-i\pi c_0^T\Big[
  \cE_\pi^{-1}G v_f ~v_f^\dagger\cF_\pi v_g ~v_g^\dagger G\cE_\pi^{-T}
+
 \frac{1}{1-e^{-i 2\pi \nu_c}} \times
\nonumber\\
&\times
\left[
 -
    \cE_\pi^{-1}G v_f ~v_f^\dagger\cE_\pi (1-R) v_d ~v_d^\dagger G\cE_\pi^{-T}
 +
    \cE_0^{-1}G v_c ~v_c^\dagger G(1-R)\cE_0\cE_0^{-T}
 -
  \cE_0^{-1}G (1-R) v_c   ~v_c^\dagger G\cE_0^{-T}
  \right]
\Big] d_0
\nonumber\\
\end{align}
where we used $\cE_\pi^{-T}-\cE_0^{-T}= -\oh (1-R) G^{-1}\cE_0
\cE_0^{-T}$ in the second term of the second line and the
orthogonality of $v_f$ and $v_c$ to substitute $\cE_\pi
\cE_0^{-1}\rightarrow  \Delta\cF \cE_0^{-1}=- \oh \cE_\pi^T(1-R)$
in the first term of the second line 
\begin{align}
=&
-i\pi c_0^T\Big[
  \cE_\pi^{-1}G v_f ~v_f^\dagger\cF_\pi v_g ~v_g^\dagger G\cE_\pi^{-T}
  +\cE_\pi^{-1}G v_f ~v_f^\dagger\cF_\pi v_d ~v_d^\dagger G\cE_\pi^{-T}
  + \cE_0^{-1}G v_c ~v_c^\dagger \cF_0\cE_0^{-T}
\Big] d_0
\nonumber\\
=&
-i\pi c_0^T\Big[ \cE_0^{-1} \cF_0 \cE_0^{-T}
\Big] d_0
\end{align}

\subsection{Consistency of the tachyonic vertexes}
Given the length of the previous computations we restrict ourselves to
show that the desired constraints are satisfied by the tachyonic
vertexes in the case $ker (\cF_\pi-\cF_0)=\emptyset$.

\subsubsection{Basic building blocks of the tachyonic vertexes and their properties}
We define now some basic build blocks of the tachyonic vertex
operators and compute some properties which turn out to be useful in
checking the wanted properties needed for defining a good CFT.
We define therefore
\begin{align}
B_0(k) &= e^{i k_i x^i_0}
\nonumber\\
B_L(k,z) &= e^{i k_i \Xln^i(z)}
\nonumber\\
B_R(k,\bar z) & =e^{i k_i \Xrn^i(\bar z)}
=e^{i (R_0^T k)_i \Xln^i(\bar z)}
\end{align}
where in the last line we used the $X$ boundary conditions.
They have the following non trivial products ($|z| > |w|$)
\begin{align}
B_0(k) ~ B_0(l)
&=
B_0(k+l) e^{-i \pi \alpha' ~ k_i [(\hF_\pi -\hF_0)^{-1}]^{i j} l_j }
\nonumber\\
B_L(k,z)~B_L(l,w)
&=
: e^{i k_i \Xln^i( z)+ i l_i \Xln^i(w)} : 
~ e^{2\alpha'~ \sum_a k^T v_a~ v_a^\dagger l ~\hat g_{-\nu_a,0}\left(\frac{w}{z}\right)}
\nonumber\\
B_R(k,\bar z)~B_R(l,\bar w)
&=
: e^{i (R_0^T k)_i \Xln^i(\bar z)+ i (R_0^T l)_i \Xln^i(\bar w)} : 
~ e^{2\alpha'~ \sum_a k^T R_0 v_a~ v_a^\dagger R_0^T l ~\hat
  g_{-\nu_a,0}\left(\frac{\bar w}{\bar z}\right)}
\nonumber\\
B_L(k,z)~B_R(l,\bar w)
&=
: e^{i k_i \Xln^i( z)+ i (R_0^T l)_i \Xln^i(\bar w)} : 
~ e^{2\alpha'~ \sum_a k^T v_a~  v_a^\dagger R_0^Tl ~\hat
  g_{-\nu_a,s}\left(\frac{\bar w}{z}\right)}
\end{align}
where 
we have explicitly written the sheet on which $g_{-\nu_a,s}$  is
defined since  $-2\pi < arg(\bar w / z) <0$ 
and we have defined the fundamental sheet $s=0$ by
$\pi < arg(\bar w / z) < \pi$ so we can have $s=0,1$.

Using the previous products it is not difficult to derive the following
properties of the commutation / analytic continuation to the range $|z| < |w|$
\begin{align}
B_0(k) ~B_0(l)
&=
B_0(l) ~B_0(k) 
~e^{-i 2 \pi \alpha' ~ k_i [(\hF_\pi -\hF_0)^{-1}]^{i j} l_j }
\nonumber\\
{}[B_L(k,z)~B_L(l,w)]_{\mbox{an.cont.}}
&=
B_L(l,w)~B_L(k,z)
~ \exp\{-2\alpha'~ \sum_a k_a l_{-a} 
\frac{\pi}{\sin \pi \nu_a} e^{i \pi \nu_a ~arg(w/z)} \}
\nonumber\\
&=
B_L(l,w)~B_L(k,z)
~ \exp\{-i 4\pi \alpha'~ k_i 
\left( \frac{R^{ \Si(-arg(w/z)) }}{\uno-R } G^{-1} \right)^{i j} l_j  \}
\nonumber\\
\COMMENTOO{\bf ...........
B_R(k,\bar z)~B_R(l,\bar w)}
&=
\nonumber\\
{}[B_L(k,z)~B_R(l,\bar w)]_{\mbox{an.cont.}}
&=
B_R(l,\bar w)~B_L(k ,z)
~ \exp\{-i 4\pi \alpha'~ k_i 
\left( \frac{R }{\uno-R } G^{-1} \right)^{i j} (R_0^Tl)_j  \}
\nonumber\\
&=
B_R(l,\bar w)~B_L(k ,z)
~ \exp\{+i 4\pi \alpha'~ (R_0^Tl)_i 
\left( \frac{\uno }{\uno-R } G^{-1} \right)^{i j} k_j  \}
\nonumber\\
\end{align}
where we have used the spectral decomposition for the matrix $R$ given
in  eq. (\ref{spectralGR}) and some attention to the range of $arg(\bar w/z)$ must be used to
get the result.

\subsubsection{The tachyonic vertexes}
We are now in the position of writing the tachyonic
vertexes for both the dipole and closed tachyons.
The tachyonic emission vertex from the $\sigma=0$ boundary reads 
\begin{align}
\cV_{T_0}(x, k)
&=
x^{ -\alpha' k_{i } \cG_0^{i j} k_{j }}
~e^{\alpha'  \Delta_c(1) ~\hat k_{c} \hat k_{-c} }
~e^{i k_{i} x^i_0}
~: e^{i k_\mu X^\mu_L(x)}:  
:e^{i k_{i } \left(\frac{\uno+R_0}{2}\right)^i_{\, j} \Xln^j(x) }  :
~\Lambda_{(0)}
\nonumber\\
&=
x^{- \alpha' k_{i } \cG_0^{i j} k_{j} }
~e^{\alpha' \Delta_c(1) ~\hat k_{c } \hat k_{-c } }
~: e^{i k_\mu X^\mu_L(x)}
~B_0(k) ~B_L( \frac{\uno+R_0^T}{2} k,x)
~\Lambda_{(0)}
\end{align}
where $\hat k_c=v_c^\dagger \cE_0^{-T} G k$
and we have written the last line to show why we
talked of basic building blocks for the $B$s.

In a similar way we can write the explicit expressions for the
emission of a dipole tachyon from the $\sigma=\pi$
boundary of the dicharged string  in a compact way as
\begin{align}
\cV_{T_\pi}(y, l)
&=
|y|^{ -\alpha' l_{i } \cG_\pi^{i j} l_{j }}
~e^{\alpha'  \Delta_c(1) ~l_{c} l_{-c} }
~e^{i l_{i} x^i_0}
~: e^{i l_\mu X^\mu_L(y)}:  
:e^{i l_{i } \left(\frac{\uno+R_\pi}{2}\right)^i_{\, j} \Xln^j(y) }  :
~\Lambda_{(\pi)}^T
\nonumber\\
&=
|y|^{- \alpha'l_{i } \cG_\pi^{i j} l_{j} }
~e^{\alpha' \Delta_c(1) ~l_{c } l_{-c } }
~: e^{i k_\mu X^\mu_L(y)}
~B_0(k) ~B_L( \frac{\uno+R_\pi^T}{2} l,y)
~\Lambda_{(\pi)}^T
\end{align}
where $l_c=v_c^\dagger \cE_\pi^{-T} G l$
and the Chan-Paton matrix is transposed because the color flows in
the opposite direction with respect to the $\sigma=0$ boundary.

Finally we can write the explicit expressions for the emission of a
closed string tachyon from the dicharged string including a cocycle
$e^{i \pi \alpha' \,k_L^T\, S\, k_L}$ (which was omitted in the main text)
as
\begin{align}
\cW_{T_c}(z,\bar z; k_L, k_R)
=&
e^{i \pi \alpha' \,k_L^T\, S\, k_L}
~: e^{i \oh k_\mu X^\mu_L(z)}:  ~: e^{i \oh k_\mu X^\mu_R(\bar z)}:  
\nonumber\\
&~
z^{ -\alpha' k_{ L i } G^{i j} k_{ L j }}
~e^{\alpha'  \Delta_c(1) ~k_{L i} v_c^i~ k_{ L\, j}  v_{-c}^j}
~e^{i k_{L i} [ (G^{-1}\cE_\pi)^i_{~j}
x^j_0 +y^i_0]}
~:e^{i k_{L~i }  \Xln^i( z) }  :
\nonumber\\
&
~{\bar z}^{ -\alpha' k_{ R i } G^{i j} k_{ R j }}
~e^{\alpha'  \Delta_c(1)   ~k_{R i} (R_0v_c)^i~ k_{ R\, j}  (R_0v_{-c})^j }
~e^{i k_{R i} [(G^{-1}\cE_\pi^T)^i_{~j}
x^j_0 + y^i_0]}
~:e^{i k_{R~i }  \Xrn^i(\bar z) }  :
\nonumber\\
=&
e^{i \pi \alpha' \,k_L^T\, S\, k_L}
~: e^{i \oh k_\mu X^\mu_L(z)}:  
~: e^{i \oh k_\mu X^\mu_R(\bar z)}:  
\nonumber\\
&~
z^{ -\alpha' k_{ L i } G^{i j} k_{ L j }}
~e^{\alpha'  \Delta_c(1) ~k_{L i} v_c^i~ k_{ L\, j}  v_{-c}^j}
~B_0(\cE_\pi^T G^{-1} k_L)\, B_L(k_L,z)
\nonumber\\
&
~{\bar z}^{ -\alpha' k_{ R i } G^{i j} k_{ R j }}
~e^{\alpha'  \Delta_c(1)   ~k_{R i} (R_0v_c)^i~ k_{ R\, j}  (R_0v_{-c})^j }
~B_0(\cE_\pi  G^{-1} k_R)\, B_R( k_R,\bar z)
\label{ClosesStrVertex}
\end{align}
where $k_L=k_R$, 
we have split the zero modes $x_0$ into left 
$x_{L 0}
=G^{-1}\cE_\pi x_0$ and 
right $x_{R 0} 
=G^{-1}\cE_\pi^T x_0$ ones and
explicitly introduced the dependence of the  ``Wilson lines'' (there
are actually no Wilson lines in non compact spaces) writing $y_0$. 
The splitting between left and right zero modes 
can be fixed by imposing that two
\ref{ClosesStrVertex} have the correct OPE as we verify in section
\ref{AppSect:twoclosed}, looking in particular to the $z$ and $\bar z$
dependent c-numbers.
\COMMENTO{
Finally we included also a purely c-number cocycle
\begin{equation}
e^{i \Phi_o(k,k)}
=
e^{i \pi (n_N A^{N M}_o n_M +m^N D_o^{N M} m^M+ m^N C^{M}_{o\,N }
  n_M )}
\end{equation}
which is necessary to reproduce the closed string OPEs but it is
trivial in the non compact case.
}

We are now ready to verify the consistency conditions for these vertexes.
\subsubsection{The open string vertexes on the opposite boundaries commute
}
The product of open string vertexes on the opposite boundaries reads 
\begin{align}
{}[\cV_{T_0}(k,x)~\cV_{T_\pi}(l,y)]_{\mbox{an.cont.}}
&=
\cV_{T_\pi}(l,y) \cV_{T_0}(k,x)
\nonumber\\
& \exp\{-i 2 \pi \alpha' ~ k^T(\hF_\pi -\hF_0)^{-1} l \}
\nonumber\\
& ~ \exp\{-i 4\pi \alpha'~ k^T \frac{\uno+R_0}{2}  
\frac{\uno }{\uno-R } G^{-1}  \frac{\uno+R_\pi^T}{2} l  \}
\end{align}
where the contribution in the second line is from $B_0$ and the one in
the third from $B_L$. We have not any contribution from Chan-Paton
factors since they acts on two different color spaces.

If we evaluate the different matrices in flat coordinates we get easily
\begin{align}
\frac{\uno+{\un R}_0}{2}
&=
\frac{\uno}{\uno + {\un   \cF}_0 }
\nonumber\\
\frac{\uno+{\un R}_\pi^T}{2}
&=
\frac{\uno}{\uno - {\un \cF}_\pi }
\nonumber\\
\frac{\uno}{\uno+{\un R}}
&=
\oh (\uno + {\un \cF}_0 )({\un \cF}_\pi-{\un \cF}_0)^{-1} (\uno - {\un \cF}_\pi )
\end{align}
and therefore it follows that the request is automatically satisfied.

\subsubsection{The open string emission vertex from $\sigma=0$ commutes with the
closed string vertex
}
The open string emission vertex from $\sigma=0$ commutes with the
closed string vertex:
\begin{align}
{}[\cV_{T_0}(l,x)~\cW_{T_c}(k_L,k_R,z,\bar z)]_{\mbox{an.cont.}}
&=
~\cW_{T_c}(k_L,k_R,z,\bar z)~ \cV_{T_0}(l,x)
\nonumber\\
& \exp\{ -i 2 \pi \alpha' 
          ~ l^T(\hF_\pi -\hF_0)^{-1} \frac{n+\hL^T m}{\sqa}   \}
\nonumber\\
& ~ \exp\{-i 4\pi \alpha'~ l^T \frac{\uno+R_0}{2}  
          \frac{\uno }{\uno-R } G^{-1}  k_L    \}
\nonumber\\
& ~ \exp\{-i 4\pi \alpha'~ l^T \frac{\uno+R_0}{2}  
          \frac{R }{\uno-R } G^{-1}  R_0 k_R   \}
\end{align}
where the second line is due to zero modes, the third to the
commutation of the open string vertex with the left moving part of the
closed one, the fourth to the commutation with the right moving part.
The terms $l^T \dots n$ automatically cancel
while the other terms $l^T \dots m$
\begin{align}
&
\exp\left\{
-i 2 \pi \sqa ~ l^T \left[ 
(\hF_\pi -\hF_0)^{-1} \hL^T
-(\hF_\pi -\hF_0)^{-1} (-B)
+ \frac{\uno+R_0}{2} \frac{\uno }{\uno-R } G^{-1}  \frac{\uno-R_\pi^T}{2}
\right] m  
\right \}
\nonumber\\
=&
\exp\left\{
-i 2 \pi \sqa ~ l^T \left[ 
(\hF_\pi -\hF_0)^{-1} (\hL^T+ \hF_\pi) 
\right] m  
\right\}
\end{align}
are not present in the non compact case.

\subsubsection{
The open string emission vertex from $\sigma=\pi$ commutes with the
closed string vertex
}
In a similar way the open string emission vertex from $\sigma=\pi$
commutes with the closed string vertex:
\begin{align}
{}[\cV_{T_\pi}(l,y)~\cW_{T_c}(k_L,k_R,z,\bar z)]_{\mbox{an.cont.}}
&=
~\cW_{T_c}(k_L,k_R,z,\bar z)~ \cV_{T_\pi}(l,y)
\nonumber\\
& \exp\{ -i 2 \pi \alpha' 
          ~ l^T(\hF_\pi -\hF_0)^{-1} \frac{n+\hL^T m}{\sqa}   \}
\nonumber\\
& ~ \exp\{-i 4\pi \alpha'~ l^T \frac{\uno+R_\pi}{2}  
          \frac{R }{\uno-R } G^{-1}  k_L    \}
\nonumber\\
& ~ \exp\{-i 4\pi \alpha'~ l^T \frac{\uno+R_\pi}{2}  
          \frac{R }{\uno-R } G^{-1}  R_0 k_R   \}
\end{align}
where the second line is due to zero modes, the third to the
commutation of the open string vertex with the left moving part of the
closed one and the fourth to the commutation with the right moving part.
The terms $l^T \dots n$ automatically cancel
while the other type of terms $l^T \dots m$ gives
\begin{align}
&
\exp\left\{
-i 2 \pi \sqa ~ l^T \left[ 
(\hF_\pi -\hF_0)^{-1} \hL^T
-(\hF_\pi -\hF_0)^{-1} (-B)
+ \frac{\uno+R_\pi}{2} \frac{R }{\uno-R } G^{-1}  \frac{\uno-R_0^T}{2}
\right] m  
\right \}
\nonumber\\
=&
\exp\left\{
-i 2 \pi \sqa ~ l^T \left[ 
(\hF_\pi -\hF_0)^{-1} (\hL^T+ \hF_0) \right] m  
\right\}
\end{align}
which again are not here for non compact case.

\subsubsection{
The product of two open string vertexes for the emission of dipole
states from the $\sigma=0$ boundary
}
The product (OPE) of two dipole tachyons must give the same result
when computed using the original vertexes or the emission vertexes
from the dicharged string.
We have already verified this property for all vertexes but it is
worth to check it directly in a simple case.
Therefore we first compute the product and 
then  we consider its OPE at the leading order.
The product reads
\begin{align}
\cV_{T_0}(x_1,k) \cV_{T_0}(x_2,l)
=&
\nonumber\\
&
\exp\Big\{ -2\alpha' \sum_a \Delta_a(1) 
\Big[
      v_{a}^\dagger G \frac{1+R_0^T}{2}k
      ~v_{-a}^\dagger G\frac{1+R_0^T}{2}l
\nonumber\\  
&~~~~~~
     + v_{a}^\dagger G \frac{1+R_0^T}{2} l  
     ~v_{-a}^\dagger G\frac{1+R_0^T}{2}k
\Big]
\Big\}
\nonumber\\
&~~~~~~
e^{\alpha' \sum_a \delta_a 
         ~v_{a}^\dagger G\frac{1+R_0^T}{2}(k+l)  
         ~v_{-a}^\dagger G\frac{1+R_0^T}{2}(k+l)  }
\nonumber\\
& \times
x_1^{\alpha' k^T \cG_{(0)}^{-1} k } ~x_2^{\alpha' l^T \cG_{(0)}^{-1} l }
\nonumber\\
& \times
\exp\left\{ -i \pi \alpha' k^T(\hF_\pi -\hF_0)^{-1} l \right\}
~B_0(k+l)
\nonumber\\
&\times
\exp\Big\{ -2\alpha' \sum_a \hat g_{-\nu_a}\left(\frac{x_2}{x_1}\right)
      v_{a}^\dagger G\frac{1+R_0^T}{2}k
      ~v_{-a}^\dagger G \frac{1+R_0^T}{2}l
\Big\}
\nonumber\\
&
:B_L(k,x_1) B_L(l,x_2):
\nonumber\\
&
\times
\Lambda_{(0)} \Lambda_{(0)} 
\end{align}
where the first line 
is the $x$ independent normalization factor,
the terms between the first and the second product symbols are due to
the $x$ dependent normalization factor,
the terms between the second and the third product symbols are due to
the zero modes,
the terms between the third and the fourth product symbols are due to
the non-zero modes
%
and the last line is due to Chan-Paton factor.
This last line is exactly the same as for the product of two vertexes 
in the dipole string.

In order to check the leading order of the OPE associated to the
previous product we expand the ``propagator'' as
\begin{align}
\hat g_{-\nu_a}\left(\frac{x_2}{x_1}\right)
=
\ln\left( 1-\frac{x_2}{x_1}\right)
+ \Delta_a(1)
+ O(x_2-x_1)^2
\nonumber\\
\end{align}
then we use the properties of $\Delta_a$ to write 
the pure c-number phase associated with the terms $k^T \dots l$
as
\begin{align}
\exp\left\{ -i \pi \alpha' k^T \left[ 
+  (\hF_\pi -\hF_0)^{-1}
+ \frac{1+R_0}{2} \frac{\uno +R }{\uno -R}  G^{-1} \frac{1+R_0^T}{2}
\right] l \right\}
\end{align}
Finally simplifying the previous phase using the $R$s definitions 
we get the right leading order
\begin{align}
\cV_{T_0}(x_1,k) \cV_{T_0}(x_2,l)
\sim&
(x_1 -x_2)^{2\alpha' k^T \cG_{(0)}^{-1]} l}
~e^{-i \pi \alpha' k^T \left( \frac{4}{\pi} D_0
        +\Theta_{(0)}
\right)        l}
~
\cV_{T_0}(x_2,k+l)
\end{align}
without any constraint.

\subsubsection{
The product of two open string vertexes for the emission of untwisted
states from the $\sigma=\pi$ boundary
}
It can be verified exactly as for the $\sigma=0$ case  done in the
previous section.
\subsubsection{
The product of two closed string vertexes
}
\label{AppSect:twoclosed}
We consider finally the leading order of  the 
product of two closed string tachyons
\begin{align}
\cW_{T_c}(k_L,k_R,z,\bar z) &~\cW_{T_c}(l_L,l_R,w,\bar w)
\sim
\nonumber\\
&
\exp\left\{ -i 2\pi \alpha' k_L^T S l_L\right\}
e^{i \Phi_o(k+l,k+l) } 
\nonumber\\
&\times
\exp\Big\{ 
-i \pi \alpha'  k_L^T \frac{\uno+R}{\uno-R} G^{-1} l_L
-i \pi \alpha'  k_R^T R_0 \frac{\uno+R}{\uno-R} G^{-1} R_0^T l_R
\Big\}
\nonumber\\
&~~~~
\exp\Big\{ 
-2 \alpha' \sum_c \Delta_c(1) k_{L}^T  G v_c~ l_L^T  G v_{ -c}
-2\alpha' \sum_c \Delta_c(1) k_{R}^T R_0 G v_c~ l_R^T R_0 G v_{ -c}
\Big\}
\nonumber\\
&~~~~
e^{\alpha' \sum_c \Delta_c(1) 
         \left[ 
         ~(k_L+l_L)^T G v_c  ~(k_L+l_L)^T G v_{-c} 
         +
         ~(k_R+l_R)^T R_0 G v_c  ~(k_R+l_R)^T R_0 G v_{-c} 
         \right]
}
\nonumber\\
&\times
w^{2 \alpha' k_L^T G^{-1} l_L} ~w^{-\Delta(k_L+l_L)}
~~~~
\bar w^{2 \alpha' k_R^T G^{-1} l_R} ~ \bar w^{-\Delta(k_R+l_R)}
\nonumber\\
&\times
\exp\Big\{ -i\pi \alpha' (\cE_\pi^T G k_L)^T (\cF_\pi-\cF_0)^{-1} (\cE_\pi^T G l_L) \Big\}
~B_0(\cE_\pi^T G (k_L+l_L))
\nonumber\\
&~~~
\exp\Big\{ -i\pi \alpha' (\cE_\pi G k_R)^T (\cF_\pi-\cF_0)^{-1}
(\cE_\pi G l_R) \Big\}
~B_0(\cE_\pi G (k_R+l_R))
\nonumber\\
&\times
\exp\Big\{
2\alpha' k_L^T G^{-1} l_L ~\ln\left(1-\frac{w}{z}\right)
+2 \alpha' \sum_c \Delta_c(1)~ 
k_{L}^T G v_c~ l_L^T G v_{ -c}
\Big\}
\nonumber\\
&~~~ B_L(k_L+l_L,w)
\nonumber\\
&\times
\exp\Big\{
2\alpha' k^T_R G^{-1} l_R ~\ln\left(1-\frac{\bar w}{\bar z}\right)
+2 \alpha' \sum_c \Delta_c(1)~ 
k_{R}^T R_0 G v_c~ l_R^T R_0 G v_{ -c}
\Big\}
\nonumber\\
&~~~ B_R(k_R+l_R,\bar w)
\nonumber\\
&\times
\exp\Big\{
-i 2 \pi \alpha' (\cE_\pi G k_R)^T (\cF_\pi-\cF_0)^{-1} (\cE_\pi^T G l_L)
-i 4 \pi\alpha' k_R^T R_0 \frac{\uno}{\uno-R} G^{-1} l_L
\Big\}
\end{align}
where the first line is from the c-number cocycle,
the terms between the first and the second product symbols are due to
the $z,\bar z$ independent normalization factor,
the terms between the second and the third product symbols are due to
the $z, \bar z$ dependent normalization factor,
the terms between the third and the fourth product symbols are due to
the zero modes,
the terms between the fourth and the fifth product symbols are due to
the non-zero modes and depend on $w$,
the terms between the fifth and the sixth product symbols are due to
the non-zero modes with a $\bar w$ dependence.
Finally the last line is the due to the commutation of the $k_R$ terms
with the $l_L$ ones.

The previous equation can then be rewritten in a way which makes clear
what constraints we get as
\begin{align}
\cW_{T_c}(k_L,k_R,z,\bar z) &~\cW_{T_c}(l_L,l_R,w,\bar w)
\sim
\nonumber\\
&
\exp\left\{ -i 2\pi \alpha' k_L^T S l_L\right\}
\nonumber\\
&\times
\exp\Big\{ -i \pi \alpha' \Big[
k_L^T \frac{\uno+R}{\uno-R} G^{-1} l_L
+k_R^T R_0 \frac{\uno+R}{\uno-R} G^{-1} R_0^T l_R
+4 k_R^T R_0 \frac{\uno}{\uno-R} G^{-1} l_L
\Big]
\Big\}
\nonumber\\
&\times
\exp\Big\{ 
-i\pi \alpha' (\cE_\pi^T G k_L)^T (\cF_\pi-\cF_0)^{-1} (\cE_\pi^T G l_L) 
-i\pi \alpha' (\cE_\pi G k_R)^T (\cF_\pi-\cF_0)^{-1} (\cE_\pi G l_R)
\Big\}
\nonumber\\
&~~~
\exp\Big\{
-2i\pi \alpha' (\cE_\pi G k_R)^T (\cF_\pi-\cF_0)^{-1} (\cE_\pi^T G l_L)
\Big\}
\nonumber\\
&\times
(z-w)^{2\alpha' k^T_L G^{-1} l_L}
~(\bar z-\bar w)^{2\alpha' k^T_R G^{-1} l_R}
~
\cW(k_L+l_L,k_R+l_R,w,\bar w)
\end{align}

We can now compare with the closed string OPEs which in the non
compact case with $k_L=k_R$ and $l_L=l_R$ reads
\begin{align}
\cW_{T_c}(z,\bar z,k_L,k_R) &~\cW_{T_c}(,w,\bar w,l_L,l_R)
\sim
\nonumber\\
& 
(z-w)^{2\alpha' k^T_L G^{-1} l_L}
~(\bar z-\bar w)^{2\alpha' k^T_R G^{-1} l_R}
~
\cW(w,\bar w,k_L+l_L,k_R+l_R)
\end{align}
Simplifying the phase we get the following constraint:
\begin{align}
&\exp\Big \{ -i 2 \pi \alpha' k_L^T \Big[
S -G^{-1}
\Big] l_L \Big\}
=
1
\end{align}
from which we fix the matrix  $S$ which enters the cocycle to be
$S=G^{-1}$.


\end{document}